\documentclass[a4paper,11pt]{article}
\pdfoutput=1 

\usepackage{jheppub} 

\usepackage{amsmath}
\usepackage{bbm}
\usepackage{amsfonts}
\usepackage{amssymb}

\usepackage{mathrsfs}

\newcommand{\be}{\begin{equation}}
\newcommand{\ee}{\end{equation}}
\newcommand{\ba}{\begin{array}}
\newcommand{\ea}{\end{array}}
\newcommand{\bea}{\begin{eqnarray}}
\newcommand{\eea}{\end{eqnarray}}

\usepackage{ulem,fancyvrb}
\usepackage{xcolor}

\title{Two-mediator dark matter models and cosmic electron excess}

\author[a]{Xuewen Liu,} 
\author[a,b,c]{Zuowei Liu}
\author[a]{and Yushan Su} 

\affiliation[a]{Department of Physics, Nanjing University, Nanjing 210093, China} 
\affiliation[b]{Center for High Energy Physics, Peking University, Beijing 100871, China} 
\affiliation[c]{CAS Center for Excellence in Particle Physics, Beijing 100049, China}

\emailAdd{xuewenliu@nju.edu.cn} 
\emailAdd{zuoweiliu@nju.edu.cn} 
\emailAdd{151150045@smail.nju.edu.cn}

\abstract{The cosmic electron energy spectrum recently observed 
by the DAMPE experiment 
exhibits two interesting features, including 
a break around 0.9 TeV and a sharp resonance near 1.4 TeV. 
In this analysis, we propose a dark matter explanation 
to both exotic features seen by DAMPE. 
In our model, dark matter annihilates in the galaxy 
via two different channels that lead to both a narrow 
resonance spectrum near 1.4 TeV and  
electron excess events over an extended energy range 
thus generating 
the break structure around TeV. 
The two annihilation channels are mediated by two 
gauge bosons that interact both with dark matter 
and with the standard model fermions. 
Dark matter annihilations through the s-channel process 
mediated by the heavier boson produce monoenergetic 
electron-positron pairs leading to the resonance excess. 
The lighter boson has a mass smaller than the dark matter such that 
they can be on-shell produced in dark matter annihilations in the galaxy; 
the lighter bosons in the final state subsequently decay to generate the extended excess events 
due to the smeared electron energy spectrum in this process.  
We further analyze constraints from various experiments, including 
HESS, Fermi, AMS, and LHC, to the parameter space of the model where 
both excess events can be accounted for. 
In order to interpret the two new features in the DAMPE data, 
dark matter annihilation cross sections in the current galaxy 
are typically much larger than the 
canonical thermal cross section needed for the correct dark matter 
relic abundance. 
This discrepancy, however, is remedied by the nonperturbative Sommerfeld 
enhancement because of the existence of a lighter mediator 
in the model. 
}

\begin{document}
\maketitle
\flushbottom


\section{Introduction}

Recently, DAMPE collaboration reported new measurements of the cosmic 
electron flux, 
which exhibit two exotic features in the energy spectrum, 
including a so-called break structure around 0.9 TeV 
and a sharp resonance near 1.4 TeV \cite{Ambrosi:2017wek}. 
The morphology of the excess electron events near 1.4 TeV in the DAMPE data hints   
a nearby cosmic ray source; a number of papers have appeared to 
interpret the DAMPE narrow resonance near 1.4 TeV, 
including astrophysical sources  
\cite{Yuan:2017ysv, Fang:2017tvj,Huang:2017egk,Cholis:2017ccs}
and dark matter (DM) sources \cite{Liu:2017rgs, 
Yuan:2017ysv,Fan:2017sor,Duan:2017pkq,Gu:2017gle,Huang:2017egk,
Zu:2017dzm,Tang:2017lfb,Chao:2017yjg,Athron:2017drj,
Cao:2017ydw,Duan:2017qwj,Gu:2017bdw,Chao:2017emq,Chen:2017tva,Li:2017tmd,
Zhu:2017tvk,Gu:2017lir,Nomura:2017ohi,Ghorbani:2017cey,Cao:2017sju,
Niu:2017hqe,Yang:2017cjm,Ding:2017jdr,Liu:2017obm,Ge:2017tkd,Zhao:2017nrt,
Sui:2017qra,Okada:2017pgr,Cao:2017rjr,Han:2017ars,Niu:2017lts,Nomura:2018jkd,
Yuan:2018rys,Pan:2018lhc,Wang:2018pcc}. 
So far, most papers interpret only the 1.4 TeV excess as signals 
of a local source. 
In this analysis, we argue that both the 0.9 TeV break and the 1.4 TeV resonance 
in the DAMPE electron spectrum could have a common origin.

We propose a dark matter explanation 
to both the break and the resonance in the DAMPE electron data. 
A dark matter subhalo {(SH)}
is assumed to exist in the vicinity of the solar system 
motivated by the morphology of the 1.4 TeV resonance excess. 
We assume a simple cosmic electron background, 
a single power-law form with only two parameters. 
Thus, both the break and the sharp resonance in the DAMPE 
data have emerged as results of excess electrons in 
dark matter annihilations in the nearby subhalo. 

We propose a two-mediator dark matter model (2MDM) to 
interpret the excess electrons seen by DAMPE. 
In our model, dark matter  
can annihilate in the galaxy via two different annihilation channels due to the two mediators 
that interact both with dark matter and with the standard model (SM) sector. 
The two annihilation channels produce distinct signatures in cosmic electron flux 
because of different mass hierarchies between DM and mediators. 
One of the two mediators, denoted as $V_1$, has a mass nearly 
twice of the DM particle; 
thus the dominated annihilation channel mediated by the $V_1$ boson is the 
$\chi \bar \chi \to V_1 \to e^- e^+$ process which produces   
cosmic electrons (and positrons) with energy equal to the DM mass. 
This then leads to a sharp resonance in the energy spectrum, 
when DM annihilates in a nearby subhalo. 
Because the sharp resonance in the DAMPE data occurs 
around 1.4 TeV, we take the DM mass to be 1.5 TeV.

The other mediator, denoted by $V_2$, is much lighter than DM 
such that the pair-production of on-shell $V_2$ bosons 
($\chi \bar \chi \to V_2 V_2$) becomes 
the primary DM annihilation channel among the processes mediated 
by the $V_2$ boson. 
The lighter mediator $V_2$ in the final state further decays 
to produce SM fermions. 
If the $V_2$ boson can directly decay into a pair of electron and positron, 
the electrons (and positrons) have a box-shape energy spectrum 
which is centered at one-half of the DM mass 
and has a width determined by the mass ratio between $V_2$ and DM. 
The box-shape electron energy spectrum is further altered  
 during the propagation between the source 
(the DM subhalo) and the observation point (the DAMPE satellite) to 
generate an extended excess in the electron 
energy spectrum. 
This then gives rise to a ``break'' structure roughly 
at one-half of the DM mass ($\sim 750$ GeV in our case) 
in the electron energy spectrum observed by DAMPE. 
In our model, because the $V_2$ boson is $L_\mu - L_\tau$ gauged so that 
the electrons originating from $V_2$ decays only carry a fraction of the total energy, 
the electron energy spectrum is further smeared.

Thus, both the break and the sharp resonance observed by DAMPE 
arise due to the electrons coming from DM annihilations in the  
2MDM model. 
We begin in section \ref{sec:model} by presenting 
the two-mediator DM models in which 
DM interacts with SM via two different gauge bosons. 
In section \ref{sec:BG}, 
we provide the cosmic electron background 
used in the analysis. 
In section \ref{sec:flux},  
we describe the method to compute electron flux 
from DM annihilations both in the DM subhalo 
and in the Milky Way (MW) halo, 
as well as the method to compare our calculations with 
DAMPE data. 
In section \ref{sec:DAMPE}, 
we compute the DAMPE signals expected in the 
two-mediator dark matter models. 
In section \ref{sec:HESS}, 
we analyze HESS constraints on our DM model. 
In section \ref{sec:Fermi}, 
we study the constraints from the Fermi isotropic gamma ray background measurements. 
In section \ref{sec:AMS}, 
we investigate AMS constraints on our DM model. 
In section \ref{sec:ATLAS}, 
we calculate ATLAS constraints on our DM model. 
In section \ref{sec:RD}, 
we discuss Sommerfeld enhancement in our model 
and the impacts on DM relic abundance.   
In section \ref{sec:sum}, 
we summarize our findings.


\section{The two-mediator dark matter model}
\label{sec:model}

\begin{figure}[!htbp]
\begin{centering}
\includegraphics[width=0.65\columnwidth]{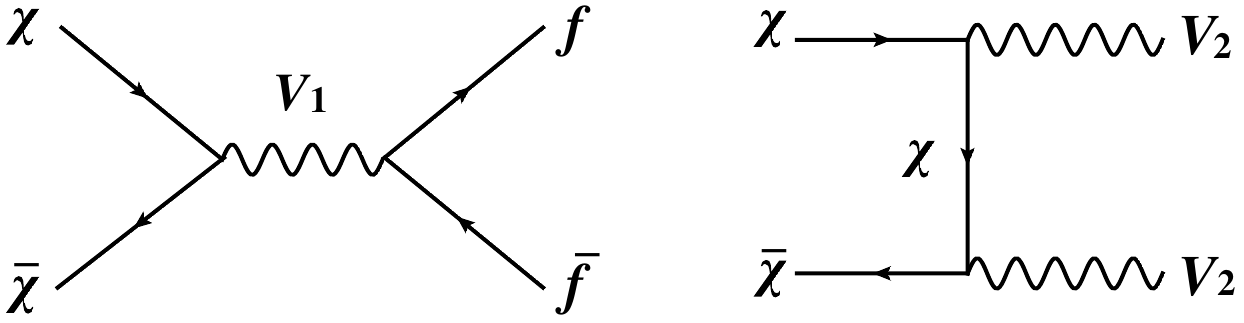}
\caption{Feynman diagrams for the two different annihilation channels 
with different mediators.}
\label{fig-diagram}
\end{centering}
\end{figure}

We consider DM models in which DM is a Dirac fermion and 
charged under two $U(1)$ fields; 
the corresponding gauge bosons are $V_1$ and $V_2$. 
Hinted by the sharp resonance near 1.4 TeV 
in the DAMPE data, we fix the DM mass at 1.5 TeV. 
The mass of the $V_1$ boson can be near 3 TeV; 
The mass of the $V_2$ is lighter than the DM mass 
such that $V_2$ can be on-shell produced in DM 
annihilations in the DM halo. 
Thus the relevant DM annihilation channels are 
\bea
&&\chi \chi \to V_1 \to f \bar f \\
&&\chi \chi \to V_2 V_2 \to  f \bar f f' \bar f'
\eea
The Feynman diagrams for the two annihilation channels 
are shown in Fig.\ (\ref{fig-diagram}). 

In our analysis, we consider two cases: 
(1) $V_1$ is electrophilic and 
$V_2$ is the $L_\mu - L_\tau$ gauged; 
(2) $V_1$ is hidden and kinetically mixes with the SM hypercharge 
and $V_2$ is $L_\mu - L_\tau$ gauged. 
Below we provide one concrete model in the latter case where  
a Dirac fermion dark matter particle $\chi$ couples to two spin-$1$ 
mediators $V_{1}$ and $V_{2}$:
\bea\label{eq:eff_lag}
\mathcal{L} &\supset  &-\frac14 V_1^{\mu\nu}V_{1\mu\nu}  + g_{1 }\bar{\chi} 
\gamma_\mu \chi V_1^\mu + \frac\epsilon 2 V_{1}^{\mu\nu}B_{\mu\nu} 
- \frac14 V_2^{\mu\nu}V_{2\mu\nu} + g_{2 }\bar{\chi} \gamma_\mu \chi V_2^\mu 
\nonumber \\ 
&+& 
g_2 (\bar{\mu} \gamma_\mu \mu -\bar{\tau} \gamma_\mu \tau +\bar{\nu}_\mu 
\gamma_\mu P_L \nu_\mu -\bar{\nu}_\tau \gamma_\mu P_L  \nu_\tau) V_2^\mu, 
\eea
where $\epsilon$ is the kinetic mixing parameter between gauge boson $V_{1\mu}$ 
and the SM $U(1)_Y$ hypercharge gauge boson $B_{\mu}$, 
$g_1 (g_2)$ is the gauge coupling for bosons $V_{1\mu}$ ($V_{2\mu}$). 
Here DM fields carry the $L_\mu - L_\tau$ quantum number. 
$V_{1\mu\nu} (V_{2\mu\nu})$ is the field strength of $V_{1\mu}$ ($V_{2\mu}$); 
$B_{\mu\nu} $ is the SM $U(1)_Y$ field strength.


\section{Cosmic electron background}
\label{sec:BG}

Because DAMPE does not distinguish the charge of 
electron/positron events, 
we assume the following single power-law background (BG) for the total flux 
of electron and positron, 
\be 
\Phi^{\text{BG}}_{e^\pm}  = C E^{-\gamma}, 
\label{eq:bg} 
\ee
where $C$ and $\gamma$ are free parameters to be determined by data. 
In our analysis, we use the first eight points and the last eight points 
in the DAMPE data \cite{Ambrosi:2017wek} 
to fit this single power-law background, 
and obtain the following best-fit parameters: 
$C = 458$  (GeV m$^{2}$ s sr)$^{-1}$ 
and 
$\gamma$= 3.25. 
We use these background parameters throughout our study.  
Since DAMPE is unable to discriminate electrons from positrons, 
we will use the word ``electron'' in this paper to collectively denote both electron and positron 
when there is no confusion.

\section{Electron flux from DM annihilations}
\label{sec:flux}

The sharp resonance of the excess events near 1.4 TeV in the DAMPE data 
hits a nearby electron/positron source. 
To fit the spectrum of the DAMPE data, we introduce a nearby DM subhalo 
with an NFW density profile \cite{Navarro:1996gj}
\be
\rho(r) = \rho_s {(r/r_s)^{-\gamma} \over (1+r/r_s)^{3-\gamma}}. 
\ee
We use the following parameters for the subhalo: $\gamma=0.5$, 
$\rho_s=100~{\rm GeV/cm}^3$, 
and $r_s = 0.1~{\rm kpc}$ \cite{Liu:2017rgs}. 
The distance between the subhalo and us (denoted by $d_s$) is also crucial to 
the cosmic ray spectrum. We find that the above subhalo with 
$d_s=0.3$ kpc can fit the DAMPE data well. 
The above values of the four parameters are assumed for the 
subhalo in our analysis if not specified otherwise.

The electron/positron flux can originate from dark matter annihilations 
both in the Milky Way 
 dark matter halo and in a nearby subhalo. 
The electron flux from DM annihilations in the MW halo 
(denoted by $\Phi^{\chi-\text{MW}}$) 
is computed via PPPC4DMID \cite{Cirelli:2010xx}. 
For the electron flux arising from DM annihilations in a nearby subhalo,  
we use the Green's function method
\cite{Ginzburg, Kuhlen:2009is, Delahaye:2010ji, Liu:2017rgs} 
\be
\Phi^{\chi-{\rm SH}}({\bf x}, E) = {v_e \over 4\pi} \int d^3 x_s \int dE_s\, 
G({\bf x}, E; {\bf x}_s, E_s) Q({\bf x}_s, E_s),
\ee
where $G({\bf x}, E; {\bf x}_s, E_s)$ is the Green's function, 
$Q({\bf x}_s, E_s)$ is the source term due to DM annihilation, 
$v_e$ is the electron velocity, 
and $\Phi^{\chi-{\rm SH}}({\bf x}, E)$ 
is the electron flux due to DM annihilations 
in the subhalo, 
which has the unit of (GeV$^{-1}$ m$^{-2}$ s$^{-1}$ sr$^{-1}$).
Here the subscript $s$ indicates  
the quantities associated with the DM source. 
The Green's function can be calculated via  
$G({\bf x}, E; {\bf x}_s, E_s) = 
b(E)^{-1} (\pi \lambda^2)^{-3/2}
\exp\left[-{({\bf x-x}_s)^2/\lambda^2}\right]$  
with the propagation scale $\lambda$ being given by 
$\lambda^2 =  \int_E^{E_s} dE' {D(E') / b(E')}$ 
where the energy loss coefficient $b(E)=b_0 (E/\text{GeV})^2$ 
with $b_0=10^{-16}$ GeV/s 
and the diffusion coefficient $D(E)=D_0(E/\text{GeV})^\delta$ 
with $D_0$ = 11 pc$^2$/kyr, and $\delta=0.7$\ \cite{Cirelli:2008id}.  
The source function due to DM annihilations is 
\be
Q({\bf x_s}, E_s) = {1 \over 4} { \rho^2_\chi({\bf x_s}) \over m_\chi^2 } 
\langle \sigma v \rangle {dN \over dE_s}(E_s),  
\ee
where $m_\chi$ is the DM mass, 
$ \rho_\chi({\bf x_s})$ is the DM mass density,  
$\langle \sigma v \rangle$ is the velocity-averaged DM 
annihilation cross section,  
${dN / dE_s}$ is the electron energy spectrum per DM annihilation.

Thus, in our analysis, the total electron flux is given by 
$\Phi^{\text{th}} = \Phi^{\rm BG} + \Phi^{\chi-\text{MW}} + \Phi^{\chi-\text{SH}}$ 
where we consider three major contributions: 
the cosmic ray background, 
DM annihilations  
in the MW halo, and DM annihilations in the nearby subhalo. 
To compare our calculations with the DAMPE data, 
we further take into account the ``bin effects" 
by performing the following {computation}   
\be
\Phi_i^{\rm th} =\frac{1}{E_i^{\rm max}-
E_i^{\rm min}}\int_{E_i^{\rm min}}^{E_i^{\rm max}}  
\Phi^{\rm th}  (E) dE,
\ee
where  
$E_i^{\rm min}$ ($E_i^{\rm max}$) is the lower (upper) bound of the $i$-th bin 
in the DAMPE data. 
To fit the DAMPE data, we carry out the following $\chi^2$ analysis  
\be
\chi^2=\sum_{i}\frac{( \Phi_i^{\text{th}} - \Phi_i^{\text{exp}} )^2}{\delta_i^2},
\label{eq:chi2}
\ee
where $\Phi_i^{\text{exp}}$ ($\delta_i$)
is the electron flux (uncertainty) reported by the DAMPE experiment \cite{Ambrosi:2017wek}.

\section{{DAMPE excess events in the two-mediator dark matter models}}
\label{sec:DAMPE}

In this section, we compute the electron flux 
{expected in DAMPE} 
from DM annihilations in the 
two-mediator {DM} model. 
DM annihilations both in the subhalo and in the MW halo are considered 
in our analysis. 
We use $\langle \sigma v \rangle_1$ to denote the velocity averaged DM 
annihilation cross section for the $\chi\chi\to V_1 \to f \bar f$ process, 
which is mediated by the heavier gauge boson $V_1$; 
we use $\langle \sigma v \rangle_2$ to denote the velocity averaged DM 
annihilation cross section for the $\chi\chi\to V_2 V_2$ process 
where the lighter gauge boson $V_2$ is on-shell produced.  
The annihilation cross sections $\langle \sigma v \rangle_1$ 
and $\langle \sigma v \rangle_2$ 
are mainly responsible for the resonance and the break excess events 
in the DAMPE electron spectrum respectively.


\subsection{Electrophilic and gauged $L_\mu - L_\tau$}

\begin{figure}[!htbp]
\begin{centering}
\includegraphics[width=0.45\columnwidth]{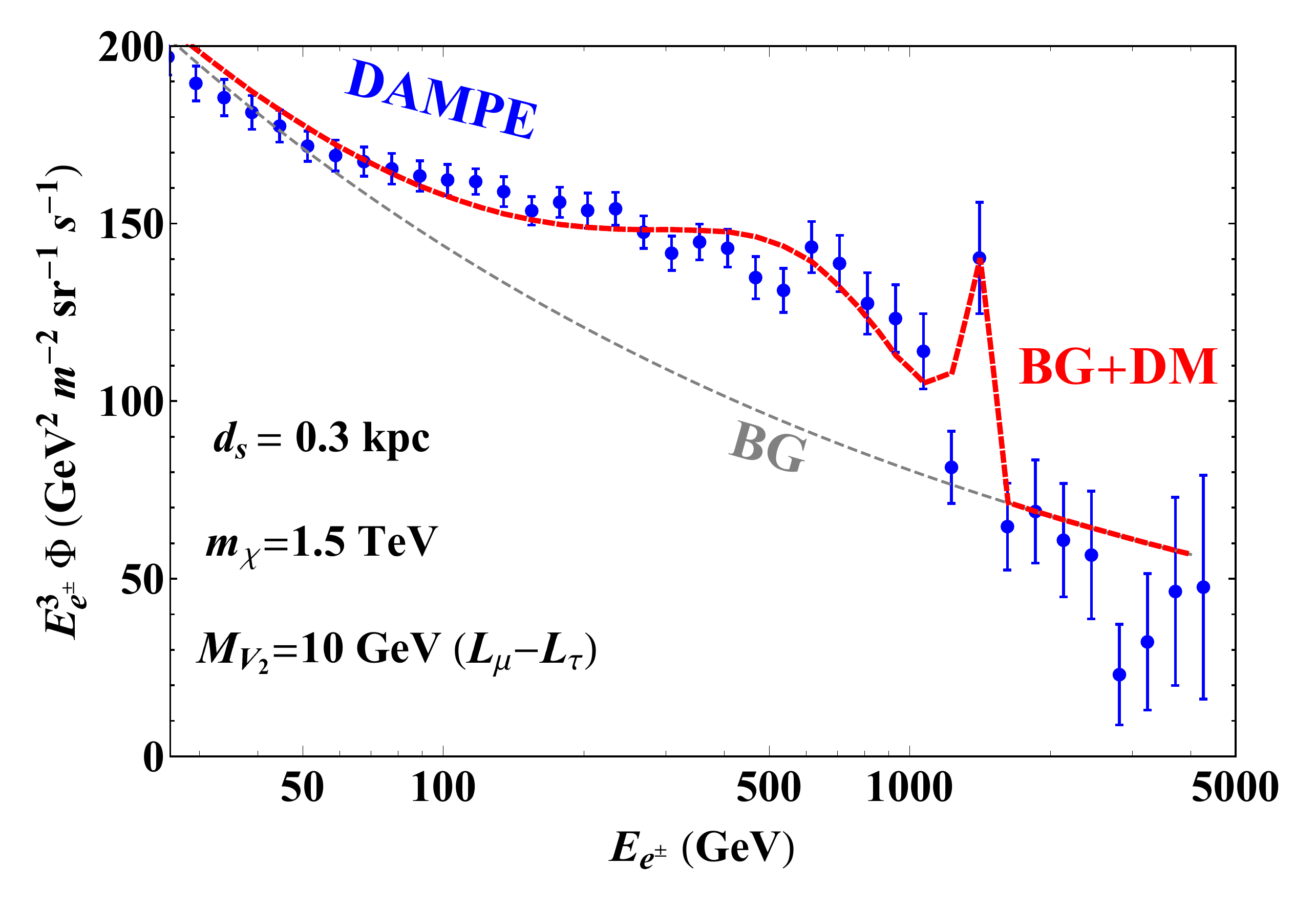}
\includegraphics[width=0.44\columnwidth]{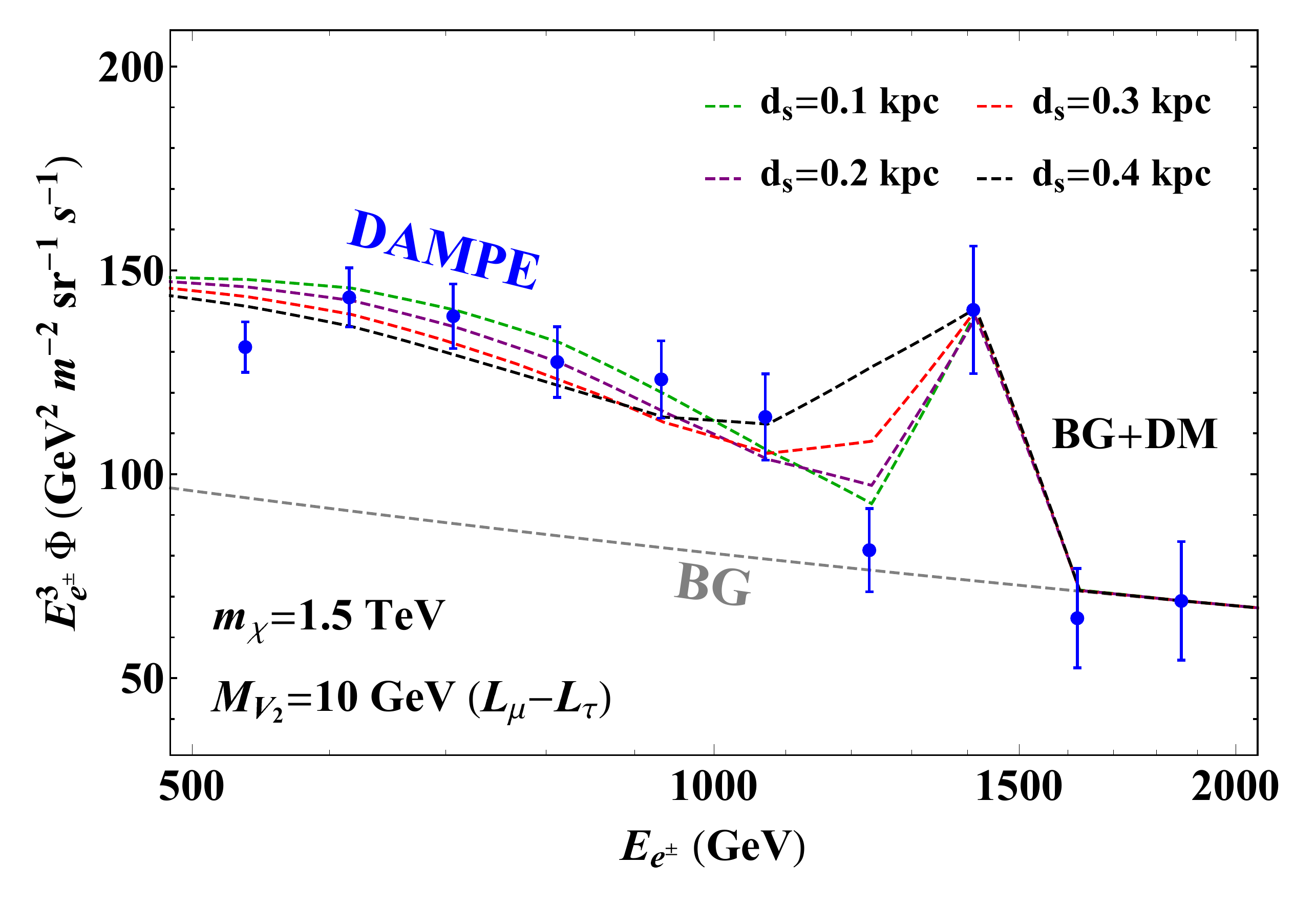} 
\caption{Left panel: 
DAMPE electron energy spectrum. Overlaid 
is the sum of cosmic ray background and the 
electron flux 
from DM annihilations from the MW halo and 
from the nearby subhalo, 
which is $d_s= 0.3$ kpc away from us. 
Right panel: 
same as the left panel except that we 
take different $d_s$ values:  
$d_s=0.1, 0.2, 0.3, 0.4$ kpc. 
Here $V_1$ is electrophilic and 
$V_2$ is $L_{\mu}-L_\tau$ gauged. 
}
\label{fig-dampe-ds2}
\end{centering}
\end{figure}

Here we first consider the two-mediator model in which the heavier mediator 
$V_1$ is electrophilic 
and the lighter mediator 
$V_2$ is the $L_\mu - L_\tau$ gauge boson. 
In this case, for the annihilation process mediated by the $V_1$, 
only $\chi\chi\to V_1 \to e^+ e^-$ can occur. 
In the annihilation processes where $V_2$ is on-shell produced, 
$V_2$ further decays into $\mu\mu$, $\tau\tau$, and $\nu\nu$ final states 
with branching ratio BR$=1/3$ for each final state. 
The energy spectrum of the $\mu\mu$ and $\tau\tau$ final states exhibits  
a box-like distribution {that is centered at $m_\chi/2$} (see e.g.\   
\cite{Mardon:2009rc, Ibarra:2012dw, Abdullah:2014lla, Cline:2014dwa, Agrawal:2014oha, Cline:2015qha} 
for early studies in the context of cosmic rays). 
{In our analysis, because we assume a simple power-law background, 
there is a wide range of electron excess events, 
extending from about 50 GeV to almost over 1 TeV, 
as shown in the left panel figure of Fig.\ (\ref{fig-dampe-ds2}). 
To generate such an extended electron excess events, 
the mass of the $V_2$ boson has to be sufficiently small 
since the width of the box-shape energy spectrum 
is given by $\sqrt{m_\chi^2 - m_{V_2}^2}$. 
In addition, the $V_2$ boson in our study is also required to decay into the $\tau\tau$ 
final state. Thus we take 10 GeV as the benchmark point for the $V_2$ boson mass,  
which is assumed throughout our analysis.}

The left panel figure of Fig.\ (\ref{fig-dampe-ds2}) 
shows the DAMPE electron flux data, 
the cosmic ray background, 
and the electron flux from {both} DM annihilations {and background}, 
where the subhalo takes its default parameters. 
Here the 1.4 TeV peak shown in Fig.\ (\ref{fig-dampe-ds2}) 
mainly comes from the $\chi\chi \to V_1\to e^- e^+$  
annihilation channel; 
the 0.9 TeV break shown in Fig.\ (\ref{fig-dampe-ds2}) 
is primarily due to the $\chi\chi \to V_2V_2$ 
annihilation channel. 
The formation of the break structure in the electron energy spectrum here 
is due to several aspects of the problem here. 
Because the mass of $V_2$ in our analysis is taken to be 10 GeV, 
the box-shape energy 
spectrum of $\mu\mu$ and $\tau\tau$ have a wide energy range 
(extending almost from zero to $2m_\chi$). 
The energy loss in charged lepton decays and cosmic ray propagation 
in addition shapes the excess electrons. 
Thus one obtains an extended distribution of the excess electrons  
with a power-law break around TeV.

\begin{table}[htbp]
\begin{centering}
\begin{tabular}{|c|c|c|c|c|} \hline  
$d_s$ (kpc)    &  
0.1                  & 
0.2                  & 
0.3                  & 
0.4                 \\ \hline           
$\sigma v (\chi\chi\to e^+e^-) $(cm$^3$/s)    & 
$7.9\times10^{-27}$  & 
$2.1\times10^{-26}$  & 
$4.9\times10^{-26}$  & 
$1.1\times10^{-25}$  \\ \hline
$\sigma v (\chi\chi\to V_2 V_2)$ (cm$^3$/s)   &
$6.5\times10^{-25}$  & 
$1.3\times10^{-24}$  & 
$2.0\times10^{-24}$  & 
$2.8\times10^{-24}$  \\ \hline
\end{tabular}
\caption{Best-fitted cross sections of the  $\chi\chi\to e^+e^-$ 
and $\chi\chi\to V_2 V_2$ processes for different $d_s$ values. 
Here $V_2$ is the $L_\mu - L_\tau$ gauge boson. 
}
\label{tab-best-fits}
\end{centering}
\end{table}

We further vary the distance between the subhalo and us 
on the right panel figure {of} Fig.\ (\ref{fig-dampe-ds2}), 
while keeping the rest of the parameters fixed for the subhalo. 
For each $d_s$ value, we find the best-fit DM annihilation 
cross sections for both channels 
{in fitting the DAMPE data}, 
which are shown in 
Table~(\ref{tab-best-fits}). 
We will adopt the case in which $d_s=0.3$ kpc as the 
benchmark model for our analysis, 
in which the DM annihilation cross sections for the two 
channels take  
$\langle \sigma v \rangle_1 = 4.9 \times 10^{-26}$ cm$^3$/s 
and 
$\langle \sigma v \rangle_2 = 2.0 \times 10^{-24}$ cm$^3$/s.
Taking into account the Sommerfeld enhancement effects 
(see section (\ref{sec:RD}) for the detailed discussions), 
we find that one should have $g_2=0.68$ in order to obtain 
$\sigma v (\chi\chi\to V_2 V_2)=2.0 \times 10^{-24}$ cm$^3$/s.


\subsection{Kinetic mixing and gauged $L_\mu - L_\tau$}

Here we consider the two-mediator model in which the heavier mediator 
$V_1$ mixes with the SM hypercharge gauge boson via 
the kinetic mixing (KM) term  
and the lighter mediator 
$V_2$ is the $L_\mu - L_\tau$ gauge boson. 
The Lagrangian of this model is given in Eq.\ (\ref{eq:eff_lag}). 
Unlike the previous case in which the $V_1$ is electrophilic, 
the annihilation process mediated by the heavier mediator 
$\chi\chi\to V_1 \to f \bar f$ now produces 
all SM fermions. 
The analysis regarding the $V_2$ boson 
is similar to that in the previous section.

Regarding the $V_1$ boson in the KM case, 
there usually are four free parameters in the calculation: 
the KM parameter {$\epsilon$}, 
the gauge coupling {$g_1$}, 
the DM mass {$m_\chi$} which is now fixed at 1.5 TeV, 
{and} the mediator mass {$m_{V_1}$} which is typically near $2m_\chi$ 
to provide a sufficient annihilation rate for the DM relic abundance. 
However, in our case, there is another lighter mediator 
which can significantly change the annihilation cross section 
mediated by $V_1$ boson, via the Sommerfeld enhancement mechanism. 
Thus, 
to correctly compute the DM annihilation cross section in the halo for the 
$\chi\chi\to V_1 \to f \bar f$ process, one has to multiply %
the annihilation cross section due to $V_1$ (see e.g.\ Ref.\ \cite{Cline:2014dwa}) 
and the Sommerfeld enhancement factor due 
to the lighter $V_2$ mediator 
{(see section (\ref{sec:RD}) for the detailed discussions)} 
in {the} model.

\begin{figure}[!htbp]
\begin{centering}
\includegraphics[width=0.45\columnwidth]{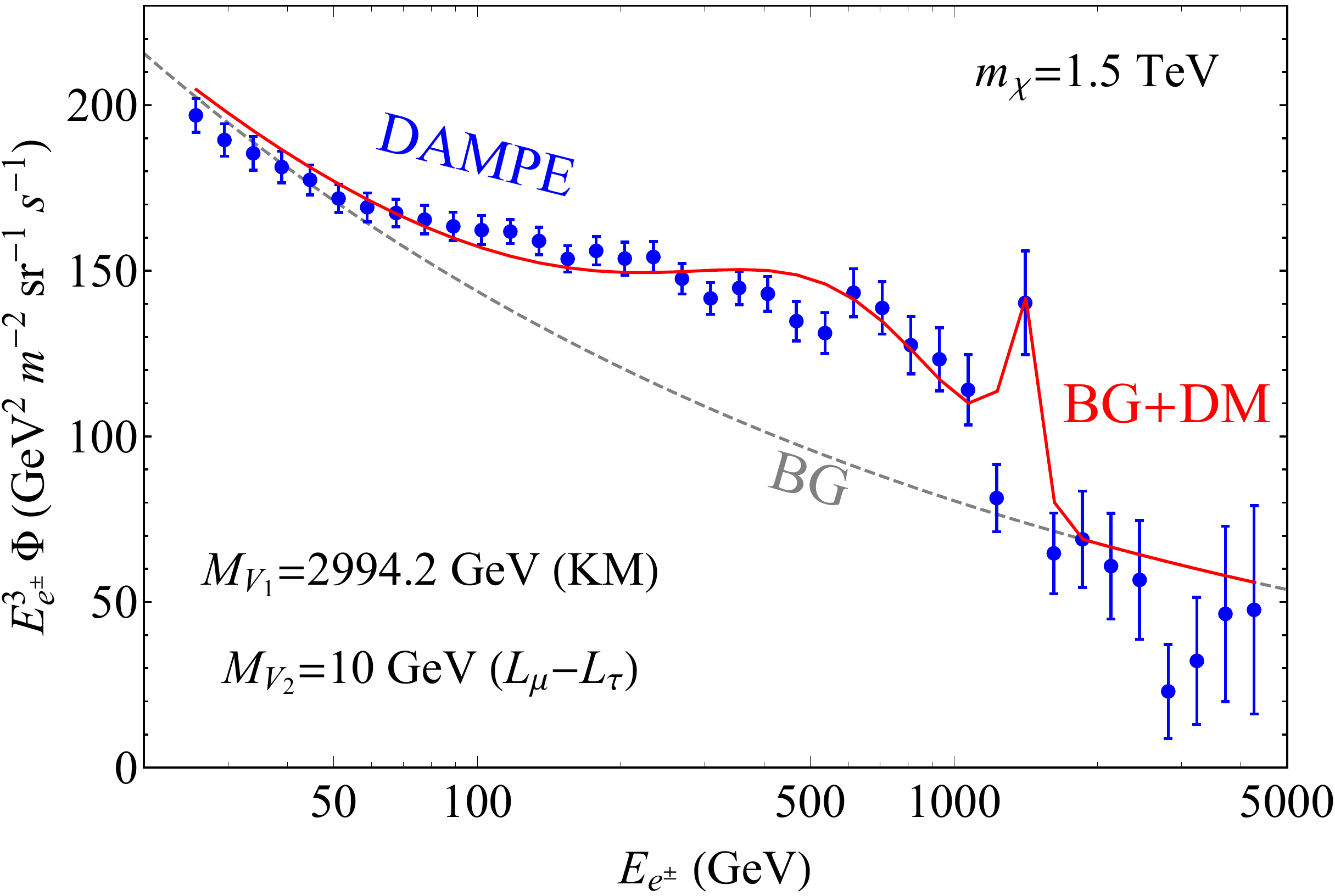}
\caption{DAMPE electron energy spectrum. Overlaid 
is the sum of cosmic ray background and the 
electron flux 
from DM annihilations from the MW halo and 
from the nearby subhalo. 
Here $V_1$ kinetically mixes with the SM hypercharge boson 
and $V_2$ is the gauge $L_{\mu}-L_\tau$ boson. 
The DM annihilation cross sections 
$\langle \sigma v \rangle_{1} =  3.9 \times 10^{-25}$ cm$^{3}$/s 
(for all SM final states in the $\chi \chi \to V_1 \to f \bar f$ process) 
and $\langle \sigma v \rangle_{2} = 2.0 \times 10^{-24}$ cm$^{3}$/s
(for the $\chi \chi \to V_2 V_2$ process) are used here. 
} 
\label{fig:DAMPE-KM}
\end{centering}
\end{figure}

In Fig.\ (\ref{fig:DAMPE-KM}), we compute the electron flux arising from the 
{2MDM} model in which the $V_1$ boson 
kinetically mixes with the SM hypercharge gauge boson, 
and the $V_2$ is the $L_\mu - L_\tau$ gauged boson. 
Here the heavier $V_1$ boson can decay into various SM fermions 
and the branching ratios are determined primarily by the hypercharge  
quantum numbers of the SM fermions. 
Since the right-handed charge lepton has a relatively large 
hypercharge, the total branching ratio of the $V_1$ boson into the three 
generation charge leptons is rather large, 
$\sum_{\ell=e,\mu,\tau}{\rm BR} (V_1 \to \ell^+\ell^-) \simeq 37\%$. 
We find that the DM annihilation cross section 
$\langle \sigma v \rangle_{1} =  3.9 \times$ 10$^{-25}$ cm$^{3}$/s 
(for all SM final states in the $\chi \chi \to V_1 \to f \bar f$ process), 
and $\langle \sigma v \rangle_{2} = 2.0 \times$ 10$^{-24}$ cm$^{3}$/s
(for the $\chi \chi \to V_2 V_2$ process) 
provide the best fit to the DAMPE data in this model. 
In Fig.\ (\ref{fig:DAMPE-KM}), 
the peak comes from the contributions of the 
$\chi \chi \to V_1 \to f \bar f$ processes 
(mainly due to the $e^+e^-$ final state), 
whereas the break are primarily due to processes mediated 
by the $L_{\mu} - L_{\tau}$ boson.

We note that the best-fit cross section for $\langle \sigma v \rangle_{2}$ 
is about the same as in the previous model so that 
$g_2 \simeq 0.68$. Taking into account the Sommerfeld enhancement factor, 
we find the model point 
($\epsilon$, $g_1$, $m_{\chi}$, $M_{V_1}$) 
= (0.01, 0.1, 1500 GeV, 2994.2 GeV)
in the parameter space can give rise to 
$\langle \sigma v \rangle_{1} =  3.9 \times 10^{-25}$ cm$^{3}$/s. 
Here the mass of the $V_1$ boson is smaller than $2m_\chi$ so that 
the invisible decay $V_1 \to \chi\chi$ cannot occur. 
In addition, the DM annihilation cross section at the early 
universe receives another suppression factor 
relative to that in the DM halo today, 
because the larger kinetic energy of the DM particles at the 
early universe moves the characteristic $\sqrt{s}$ of the DM 
annihilation process further away from the Breit-Wigner resonance 
relative to today \cite{Feldman:2008xs}. 
Because the invisible decay of the $V_1$ boson is kinetically 
disallowed here and the branching ratios into charged leptons are 
rather significant, the discovery potential of LHC for such $V_1$ boson 
is high. The discussions on LHC constraints on this model are    
given in section \ref{sec:ATLAS}.


\section{HESS constraints}
\label{sec:HESS}

The gamma ray flux produced by DM annihilations can be calculated as follows 
\begin{equation}
{d\Phi_\gamma \over dE_\gamma} = 
\sum_{i} 
{ \langle \sigma v \rangle_{i} \over 8 \pi m_{\chi}^2}
 \left( {d N_\gamma \over d E_\gamma} \right)_{i}  
J(\Delta\Omega),
\label{eq:gammaflux}
\end{equation}
where 
$m_{\chi}$ is {the} DM mass, 
$\langle \sigma v \rangle_{i}$ is the velocity-averaged DM annihilation cross section for channel $i$, 
$(d N_\gamma / d E_\gamma)_{i}$ is {the} gamma ray energy spectrum per annihilation for channel $i$, 
and $J(\Delta\Omega)$ is the J-factor for the region-of-interest (ROI).
The differential flux $d \Phi_\gamma/ dE_\gamma$ has 
unit of  (GeV cm$^2$ s)$^{-1}$.
The J-factor is computed via 
\begin{equation}
J(\Delta\Omega)=\int_{\Delta \Omega} d \Omega \int d s\, \rho^{2}_{\chi} ,
\label{eq:Jfactor}
\end{equation}
where $\Delta \Omega$ is the solid angle of the ROI, 
$\rho_{\chi}$ is the DM density,  
and $s$ is the distance along the line of light.

HESS searched for very high energy  $\gamma$-rays 
in the  inner region of the Milky Way halo, 
which is  a circular region of $1^{\circ}$ radius excluding 
a $\pm 0.3^{\circ}$ band in {the} Galactic latitude \cite{Abramowski:2011hc}
\cite{Abdallah:2016ygi}. 
With the 254-hour data accumulated \cite{Abdallah:2016ygi}, 
stringent upper bounds can be set on the DM annihilation cross sections 
for various SM final states. 
In a recent study \cite{Profumo:2017obk}, 
the HESS constraints on dark matter annihilations into on-shell mediators 
for various SM final states {are analyzed}. 
Our analysis here is similar to that in Ref.\ \cite{Profumo:2017obk}, 
but in our case, on-shell mediators annihilate into %
{a collection of} SM final states 
with branching ratios given in the %
{2MDM} model.

\subsection{HESS constraints on DM annihilations in the Galactic center}


In the following, we calculate the upper limit on the DM annihilation cross section 
$\langle \sigma v \rangle_{\chi\chi\to V_2 V_2}$ from the HESS data, 
where $V_2$ is the gauge $L_\mu - L_\tau$ boson. 
The method we use here is to rescale the limits calculated in Ref.\ \cite{Abdallah:2016ygi} 
which analyzed 254-hour data recorded by HESS. 
The details of the method can be found in Appendix (\ref{HESSlimit}).

\begin{figure}[htbp]
	\begin{centering}
		\includegraphics[width=0.45\columnwidth]{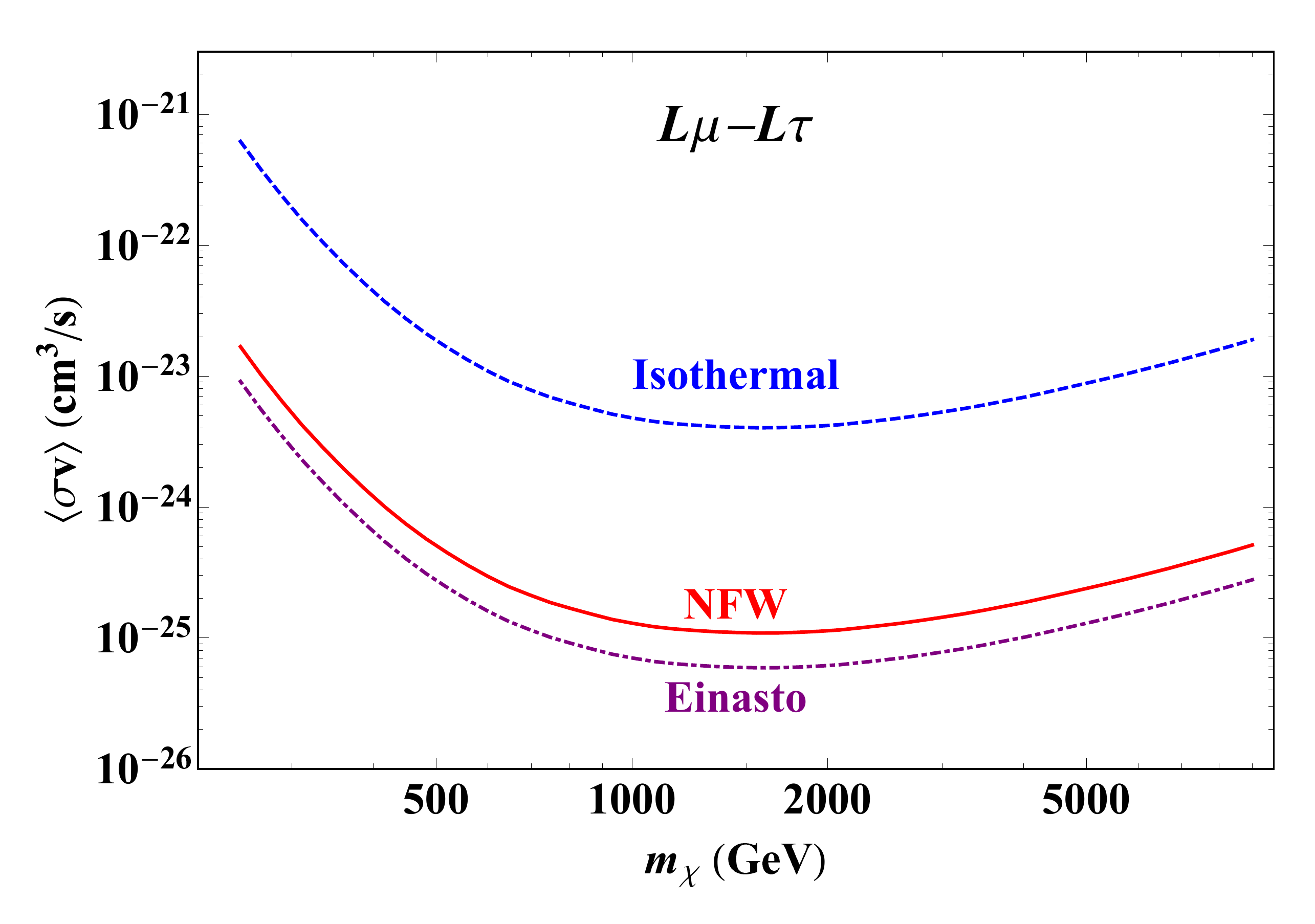}
		\caption{HESS upper limits on $\langle \sigma v \rangle (\chi\chi\to V_2 V_2)$ 
			where $V_2$ is a gauged ${L_\mu-L_\tau}$ boson. 
			We considered three different DM profiles: 
			NFW (solid), isothermal (dashed), and Einasto (dot-dashed). 
			Here only gamma rays from the MW halo are considered. 
                         The limits are computed based on the exclusion limits for the 
			$\chi\chi \to \mu^+ \mu^-$ and $\chi\chi \to \tau^+ \tau^-$ processes given in  
			Ref.\ \cite{Abdallah:2016ygi}. 
A {light} $V_2$ mass is assumed in this analysis. 
		} 
		\label{fig-hess-xw-ys}
	\end{centering}
\end{figure}

We first analyze the HESS limits on DM annihilations in the center of the galaxy. 
Because the DM distribution is not known to a good precision in the 
center of the galaxy and the gamma rays are very sensitive to the 
DM density distributions in the Galactic center, 
several DM profiles are considered in the HESS analyses 
\cite{Abramowski:2011hc} \cite{Abdallah:2016ygi}. 
We provide a comparison of the $J$-factors from different 
DM profiles in Appendix (\ref{app:Jfactor}). 
Here we consider three different DM profiles, %
NFW, Isothermal, and Einasto, to interpret the HESS constraints.

Fig.\ (\ref{fig-hess-xw-ys}) shows the 95\% CL limits on DM annihilation cross section  
$\langle \sigma v \rangle_{\chi\chi\to V_2 V_2}$ 
where $V_2$ is the gauged ${L_\mu-L_\tau}$ boson. 
For the 1.5 TeV DM annihilating into sufficiently light 
$V_2$ bosons, the HESS constraints are 
$ \langle \sigma v \rangle_2 \lesssim 
1.1\times 10^{-25}\, (4 \times 10^{-24†})$ cm$^3$/s
for the NFW (Isothermal) profile. 
Thus the DM annihilation cross section 
$ \langle \sigma v \rangle_2 =2.0\times 10^{-24}$ cm$^3$/s 
which is responsible for generating the break in the DAMPE data, 
is excluded if one considers the NFW or Einasto profile, 
but is still allowed if the isothermal profile is assumed.

\subsection{HESS constraints on the location of the subhalo}

\begin{figure}[!htbp]
\begin{centering}
\includegraphics[width=0.47\columnwidth]{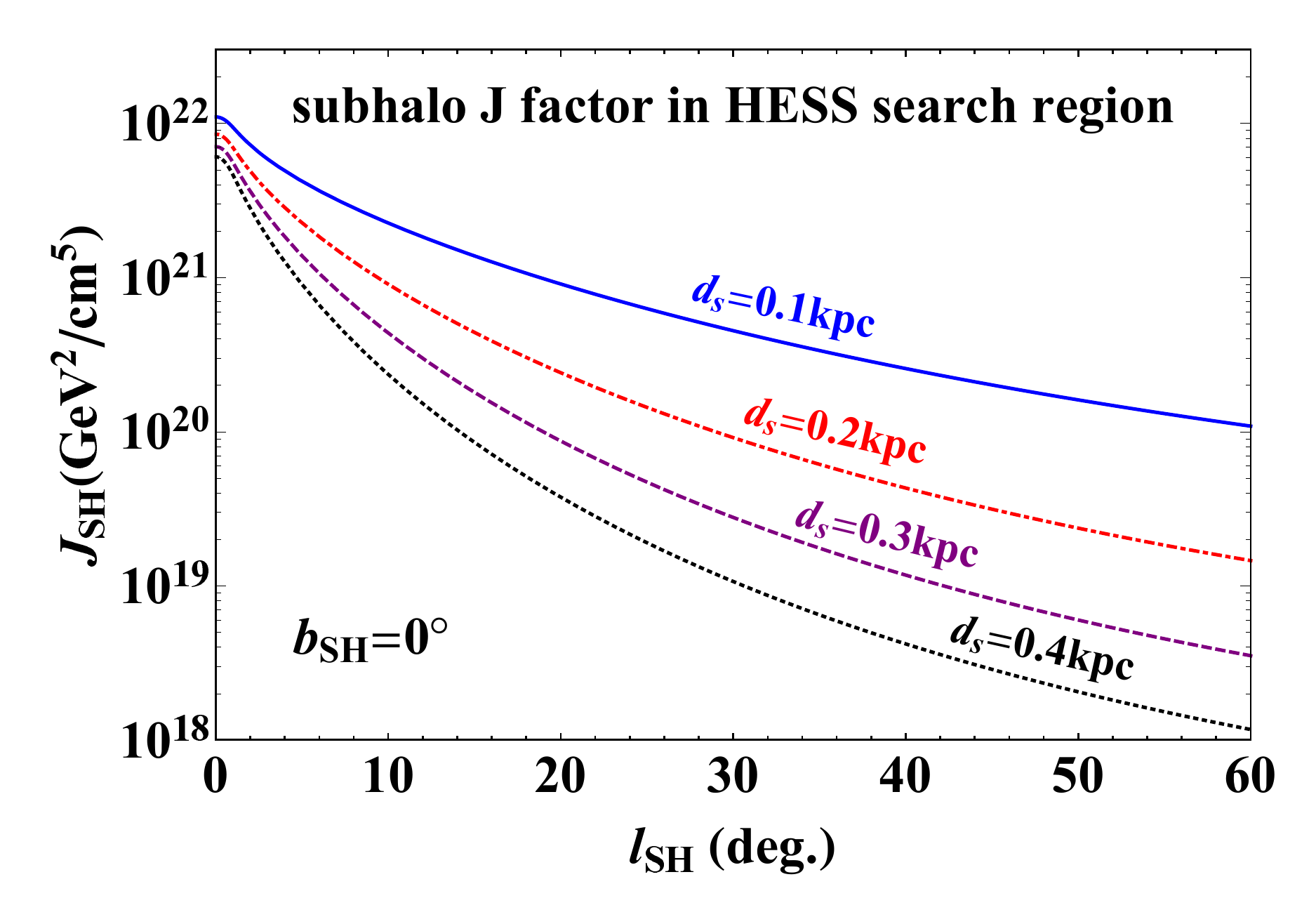}
\includegraphics[width=0.45\columnwidth]{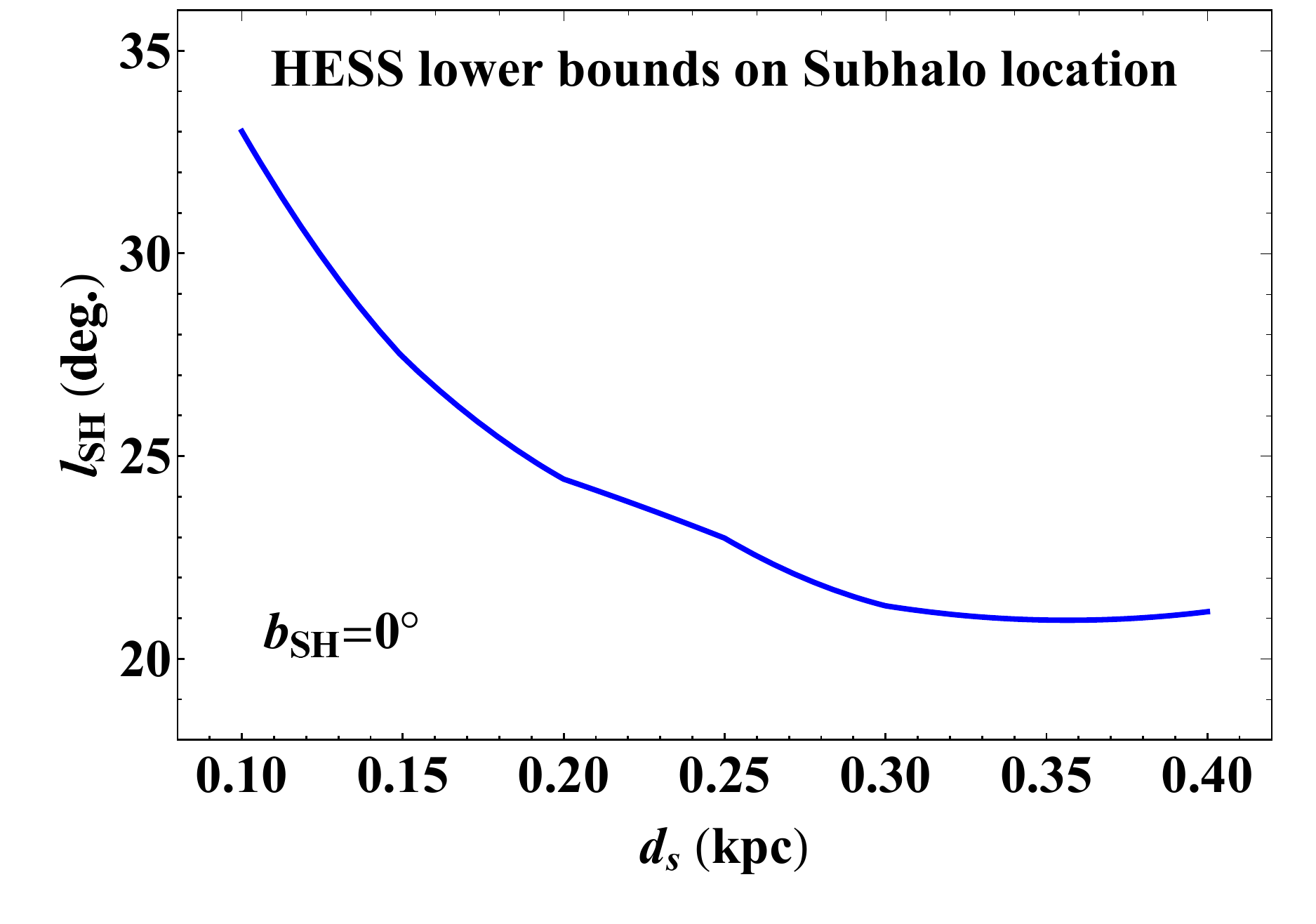}
\caption{Left panel: the subhalo $J$ factor in the HESS search region 
as the function of $l_{\rm SH}$ for different $d_s$ values.  
Right panel: HESS lower limits on $l_{\rm SH}$ as a function of $d_s$. 
Here $l_{\rm SH}$ is the angle between the 
galactic center and the center of the subhalo. 
} 
\label{fig-subhalo-to-hess-rs}
\end{centering}
\end{figure}

DM annihilations in the subhalo also contribute to  
the gamma ray flux observed by the HESS experiment. 
Because the HESS search region is $1^\circ$ around Galactic center, 
the gamma ray flux observed by HESS 
from the subhalo 
is a function of $l_{\rm SH}$ that is the angle between 
the Galactic center and the center of the subhalo. 
We compute the $J$ factor of the subhalo 
inside the HESS search region in the left panel figure 
of Fig.~\ref{fig-subhalo-to-hess-rs} 
for different $d_s$ values. 
The subhalo $J$ factor increases when the subhalo moves towards 
either the Galactic center or us.

We further determine the minimum $l_{\rm SH}$ value 
by saturating the HESS constraints on DM annihilations. 
To determine the minimum $l_{\rm SH}$ value, we use 
$
\langle \sigma v \rangle_{\rm DAMPE} 
\times ( J_{\rm MW}^{\rm iso} +J_{\rm SH} (l_{\rm SH}^{\rm min})) 
= \langle \sigma v \rangle_{\rm HESS} \times  J_{\rm MW}^{\rm iso},
$
where 
$\langle \sigma v \rangle_{\rm DAMPE}$ 
is the cross section needed for the DAMPE electron excess events, 
as given in Table\ (\ref{tab-best-fits}), 
{$J_{\rm MW}^{\rm iso}$ is the $J$ factor inside the HESS search region 
for the MW halo with the isothermal DM density profile 
which is $7.23\times10^{19}$ GeV$^2$ cm$^{-5}$, 
$\langle \sigma v \rangle_{\rm HESS}$ 
is the HESS 95\% CL upper bound on the DM annihilation cross section 
with the isothermal profile 
(which is $4 \times 10^{-24}$ cm$^3$/s as given by the isothermal curve on 
Fig.\ (\ref{fig-hess-xw-ys})),}  
$J_{\rm SH}$ is the $J$ factor inside the HESS search region 
for the subhalo. 
Because the gamma ray flux produced by the process $\chi\chi\to V_1\to e^+e^-$ 
is much smaller than $\chi\chi\to V_2 V_2$, 
we take $\langle \sigma v \rangle_{\rm DAMPE} 
\simeq \langle \sigma v \rangle (\chi\chi\to V_2 V_2)$ 
in the calculation here. 
The right panel figure of Fig.~\ref{fig-subhalo-to-hess-rs} shows the 
lower bound on the $l_{\rm SH}$ angle. 
When $d_s=0.3$ kpc, the subhalo has to be $> 21^\circ$ away from 
the Galactic center to avoid HESS constraints.

\subsection{HESS limits for both {DM} annihilation channels}

\begin{figure}[!htbp]
\begin{centering}
\includegraphics[width=0.45\columnwidth]{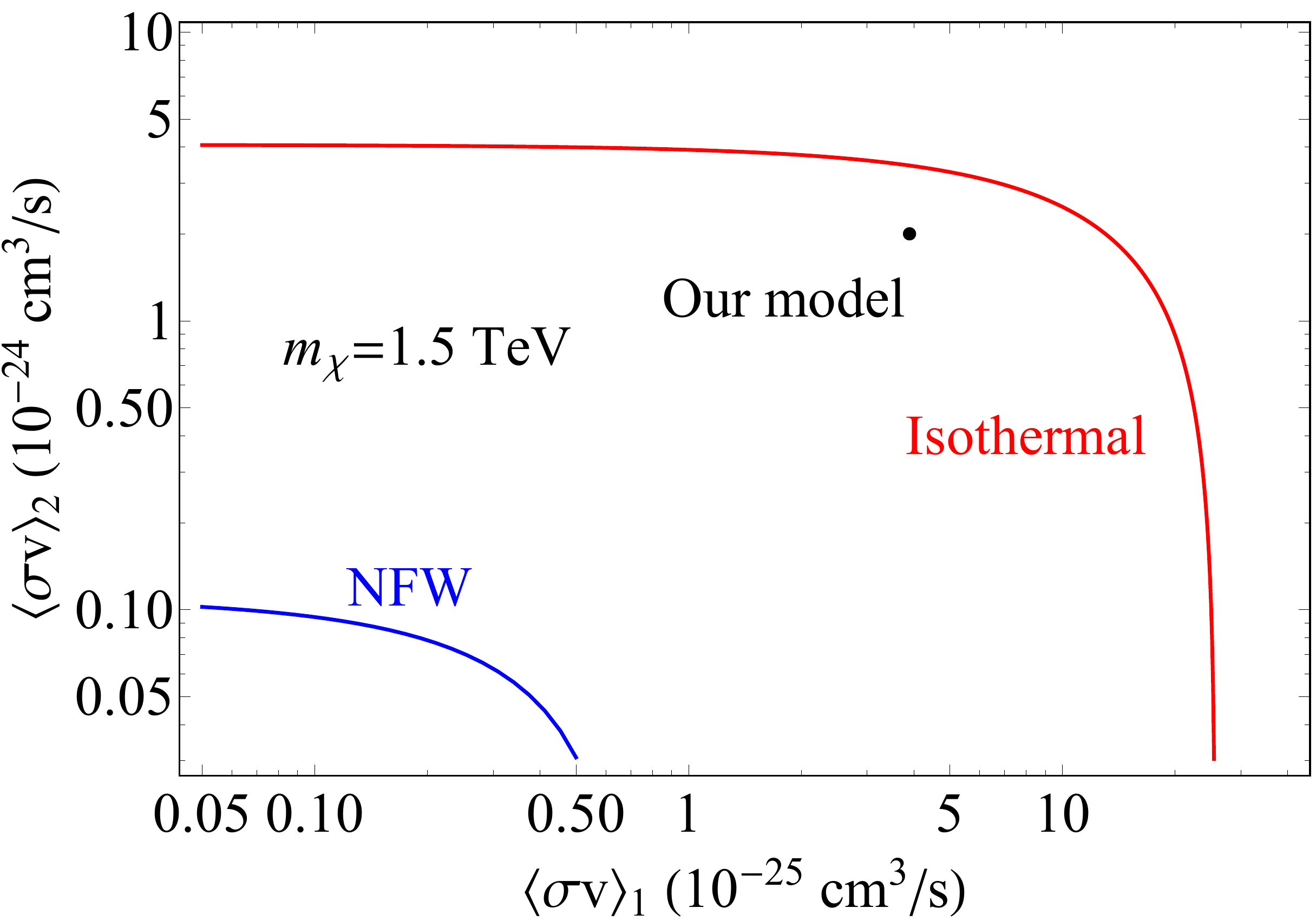}
\caption{HESS constraints on both annihilation channels. 
$\langle \sigma v \rangle_1$ is the DM annihilation 
cross section mediated by the $V_1$ boson that kinetically mixes with 
the SM hypercharge; 
{$\langle \sigma v \rangle_2$} 
is the DM annihilation 
cross section mediated by the $V_2$ boson that is 
$L_\mu - L_\tau$ gauged. Here $m_\chi$ = 1.5 TeV.  
The model point used in Fig.\ (\ref{fig:DAMPE-KM}) 
is indicated by the black point here. 
} 
\label{HESS_kineticmixing}
\end{centering}
\end{figure}

Here we analyze the HESS constraints for the model in which 
$V_1$ kinetically mixes with the SM hypercharge and 
$V_2$ is $L_\mu - L_\tau$ gauged. 
To take both channels into consideration, we use  
$
\Phi_\gamma (\langle \sigma v \rangle_1, m_{\chi}) 
+ 
\Phi_\gamma (\langle \sigma v \rangle_2, m_{\chi}) 
=
\Phi_\gamma^{95} (m_{\chi}),
$
where $\Phi_\gamma^{95}$ is the 95\% CL upper bound from the 254-h HESS data 
on the total gamma ray flux (in unit of cm$^{-2}$ s$^{-1}$) 
integrated over the energy range $160$ GeV $<E_\gamma<m_\chi$. 
Here $\Phi_\gamma (\langle \sigma v \rangle_1, m_{\chi})$ 
and 
$\Phi_\gamma (\langle \sigma v \rangle_2, m_{\chi})$ 
are the gamma rays from the two annihilation channels respectively. 
Fig.\ (\ref{HESS_kineticmixing}) shows the HESS limits 
for both $\langle \sigma v \rangle_1$ 
and 
$\langle \sigma v \rangle_2$
for the case where $m_\chi=1.5$ TeV, where 
only contributions from the MW halo are considered. 
Our model is excluded if the NFW profile is used, 
but allowed if the isothermal profile is used for the 
DM distribution in the MW halo.

\section{Fermi constraints}
\label{sec:Fermi}

Similar to the gamma ray flux measured by HESS, 
the gamma ray flux observed by Fermi due to DM annihilations is calculated as follows, 
\begin{equation}
{d \Phi_\gamma  \over dE_\gamma} = 
\sum_{i} 
{ \langle \sigma v \rangle_{i} \over 8 \pi m_{\chi}^2}
 \left( {d N_\gamma \over d E_\gamma} \right)_{i}  
\bar{J},
\label{eq:gammaflux}
\end{equation}	
where $\bar{J} = J(\Delta \Omega)/\Delta \Omega$ is the J-factor averaged over the region of interest. 
The Fermi isotropic gamma ray background (IGRB) data 
are reported as an intensity flux. 
The gamma ray flux computed in Eq.\ (\ref{eq:gammaflux}) is the intensity flux in 
unit of  (GeV cm$^2$ s sr)$^{-1}$. 

The isotropic gamma ray background measured by Fermi 
is obtained from the all-sky data excluding the $|b| < 20^{\circ}$ band 
on the Galactic plane \cite{Ackermann:2014usa}. 
The averaged $J$ factor for the Fermi isotropic gamma ray background region 
can thus be computed as follows 
\begin{equation}
\bar{J} =  \frac{\int d s \int_{|b| > 20^{\circ}} d b \, dl \cos b \, \rho^{2}_{\chi}}
{\int_{|b| > 20^{\circ}} d b \, dl \cos b},
\end{equation}	
where $\rho_{\chi}$ is the DM density, 
$b$ is the galactic latitude, 
$\ell$ is the galactic longitude, 
$s$ is the distance between the point where DM annihilates and us. 
In this study, we take into account both the MW halo and the DM subhalo 
when calculating the J-factor. In this section,  
we consider the same isothermal DM profile for the MW halo as in the HESS analysis.

\subsection{Fermi isotropic gamma ray background constraints}
\label{Fermi isotropic}

Here we compare the gamma ray flux produced by dark matter 
annihilations in the subhalo as well as in the MW halo, 
with the isotropic background measured 
by Fermi-LAT \cite{Ackermann:2014usa} to 
obtain constraints on our DM model. 
Because the galactic plane is masked in the Fermi IGRB analysis \cite{Ackermann:2014usa}, 
the constraints from Fermi IGRB are minimized when the subhalo sits on the galactic plane. 
We use $b_{\rm SH}$ to denote the galactic latitude 
{of the subhalo center.}
Thus we will set $b_{\rm SH}=0$ for our analysis unless specified otherwise.

\begin{figure}[!htbp]
\begin{centering}
\includegraphics[width=0.45\columnwidth]{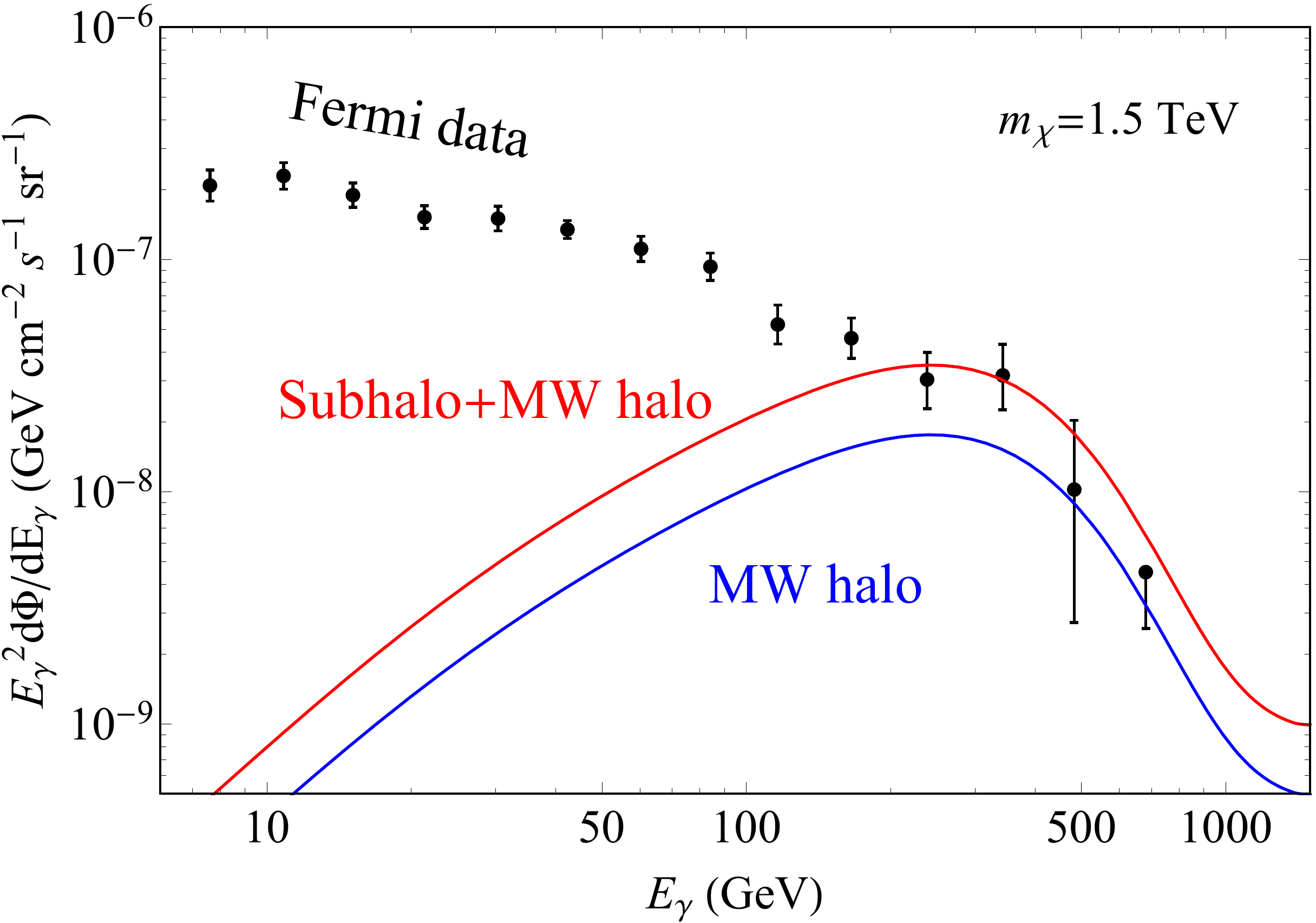}
\includegraphics[width=0.45\columnwidth]{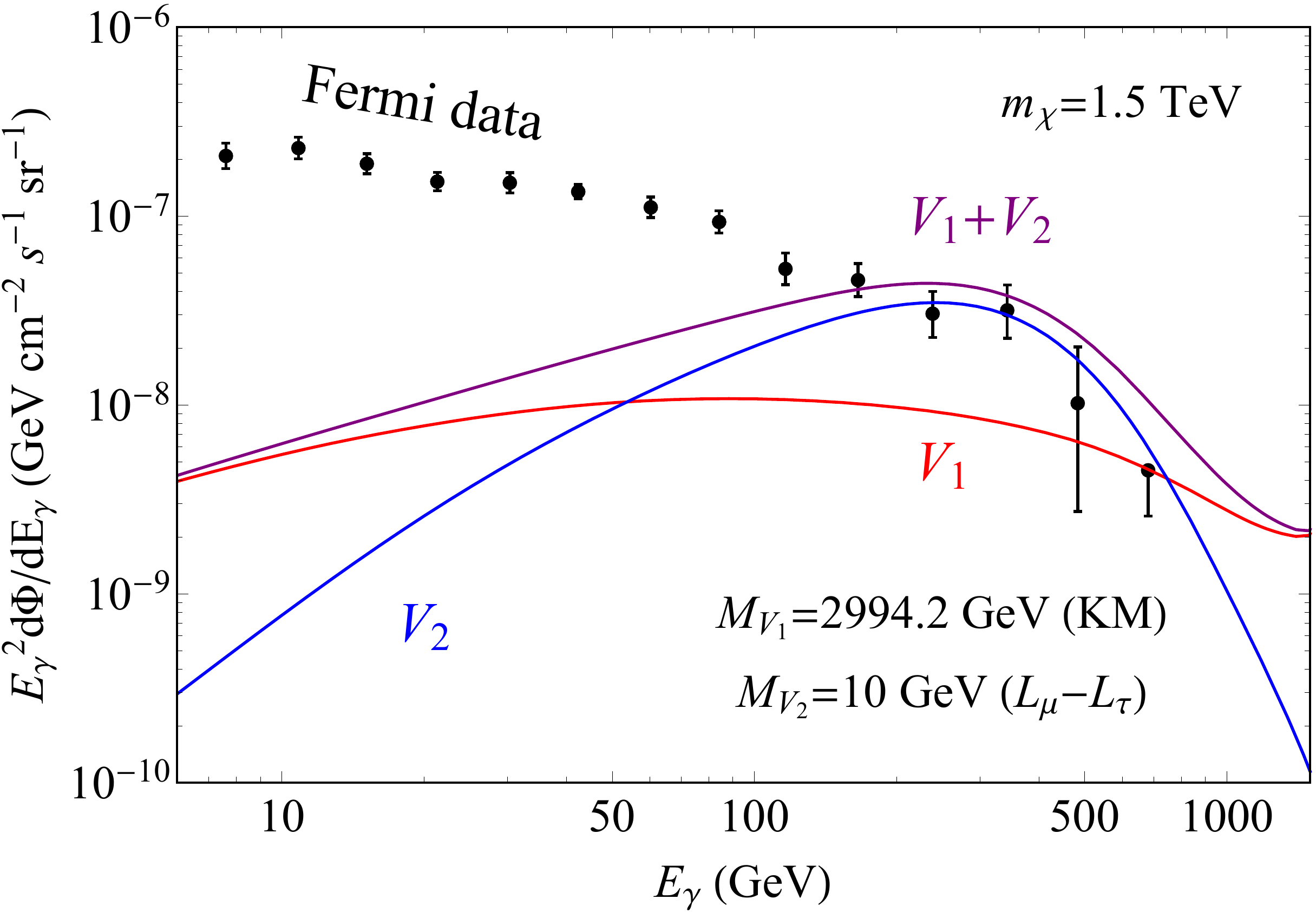}
\caption{Left panel: Fermi IGRB data \cite{Ackermann:2014usa} 
and gamma rays from DM 
where $V_1$ is electrophilic 
and $V_2$ is $L_\mu-L_\tau$ gauged. 
Right panel: 
Fermi IGRB data and gamma rays from DM 
where $V_1$ is kinetically mixed with SM hypercharge  
and $V_2$ is $L_\mu-L_\tau$ gauged and has a mass of 10 GeV. 
DM annihilation cross sections 
are the same as those in Fig.\ (\ref{fig:DAMPE-KM}).	
We use the default subhalo parameters for both figures.} 
\label{fig-subhalo}
\end{centering}
\end{figure}

The left panel figure in Fig.\ (\ref{fig-subhalo}) shows the 
Fermi IGRB data \cite{Ackermann:2014usa} 
and the gamma rays from DM annihilations for the case in which 
the heavier $V_1$ boson is electrophilic 
and the lighter $V_2$ boson is $L_\mu-L_\tau$ gauged. 
The DM annihilation cross sections for the two annihilation channels 
are ($\langle \sigma v \rangle_1$, $\langle \sigma v \rangle_2$) = 
(4.9 $\times$ 10$^{-26}$ cm$^3$/s, 2.0 $\times$ 10$^{-24}$ cm$^3$/s), 
which are the same as those in the left panel figure of 
Fig.\ (\ref{fig-dampe-ds2}). 
Here the gamma ray flux arising from the 
$\chi \chi \to e^+ e^-$ process is only about 3\% of 
that due to $\chi \chi \to V_2 V_2$ in this case.  
We find that the J-factor of the subhalo is about the same as the J-factor of the 
MW halo in the Fermi IGRB search region, 
$J_{SH} \simeq J_{MW} \simeq 6 \times 10^{21} ~ {\rm GeV^2/cm^5}$. 
We have plotted the gamma rays from the MW halo 
on the left panel figure of Fig.\ (\ref{fig-subhalo}), 
as well as the gamma rays from both the MW halo and the subhalo. 
The predicted total gamma rays in our DM model 
do not exceed {the} current Fermi IGRB bound.

For the DM model in which 
the heavier $V_1$ boson kinetically mixes with the SM hypercharge gauge boson  
and the lighter $V_2$ boson is $L_\mu-L_\tau$ gauged, 
the predicted gamma rays are shown on the right 
panel figure in Fig.\ (\ref{fig-subhalo}). 
We use the following DM annihilation cross sections 
($\langle \sigma v \rangle_1$, $\langle \sigma v \rangle_2$) = 
(3.9 $\times$ 10$^{-25}$ cm$^3$/s, 2.0 $\times$ 10$^{-24}$ cm$^3$/s)
which are the same as those in Fig.\ (\ref{fig:DAMPE-KM}). 
Unlike the DM model presented on the left panel figure of 
Fig.\ (\ref{fig:DAMPE-KM}), 
the annihilation process mediated by the $V_1$ boson on the right panel figure 
of Fig.\ (\ref{fig:DAMPE-KM}) has a larger cross section and various SM final states. 
We plotted the gamma rays from both annihilation channels 
on the right panel figure of Fig.\ (\ref{fig:DAMPE-KM}). 
We find that the isotropic gamma ray measurements  are beginning to probe this 
DM model at the high energy bins in the Fermi IGRB data.

\subsection{Fermi constraints on the subhalo}

\begin{figure}[!htbp]
\begin{centering}
\includegraphics[width=0.45\columnwidth]{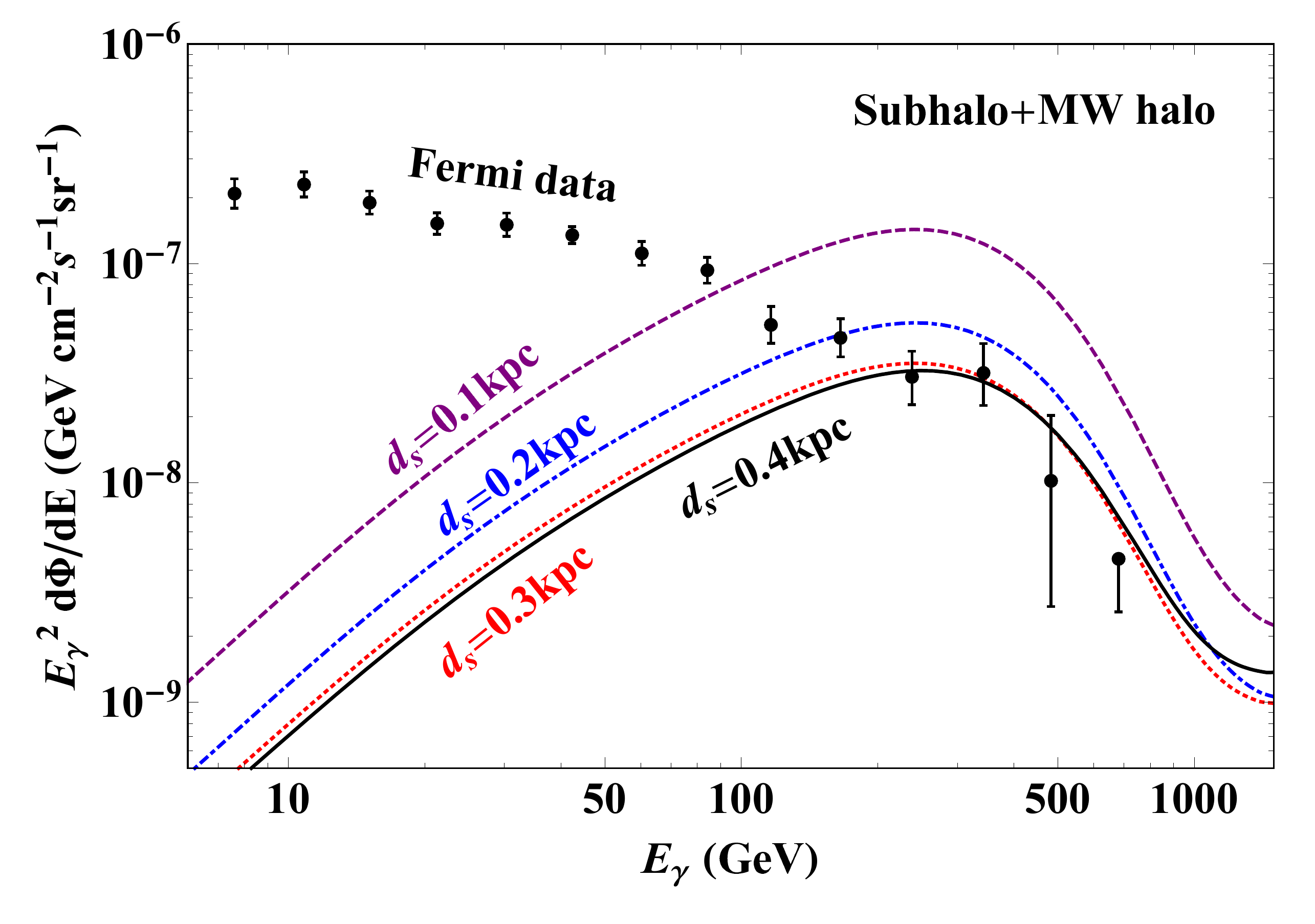}
\includegraphics[width=0.45\columnwidth]{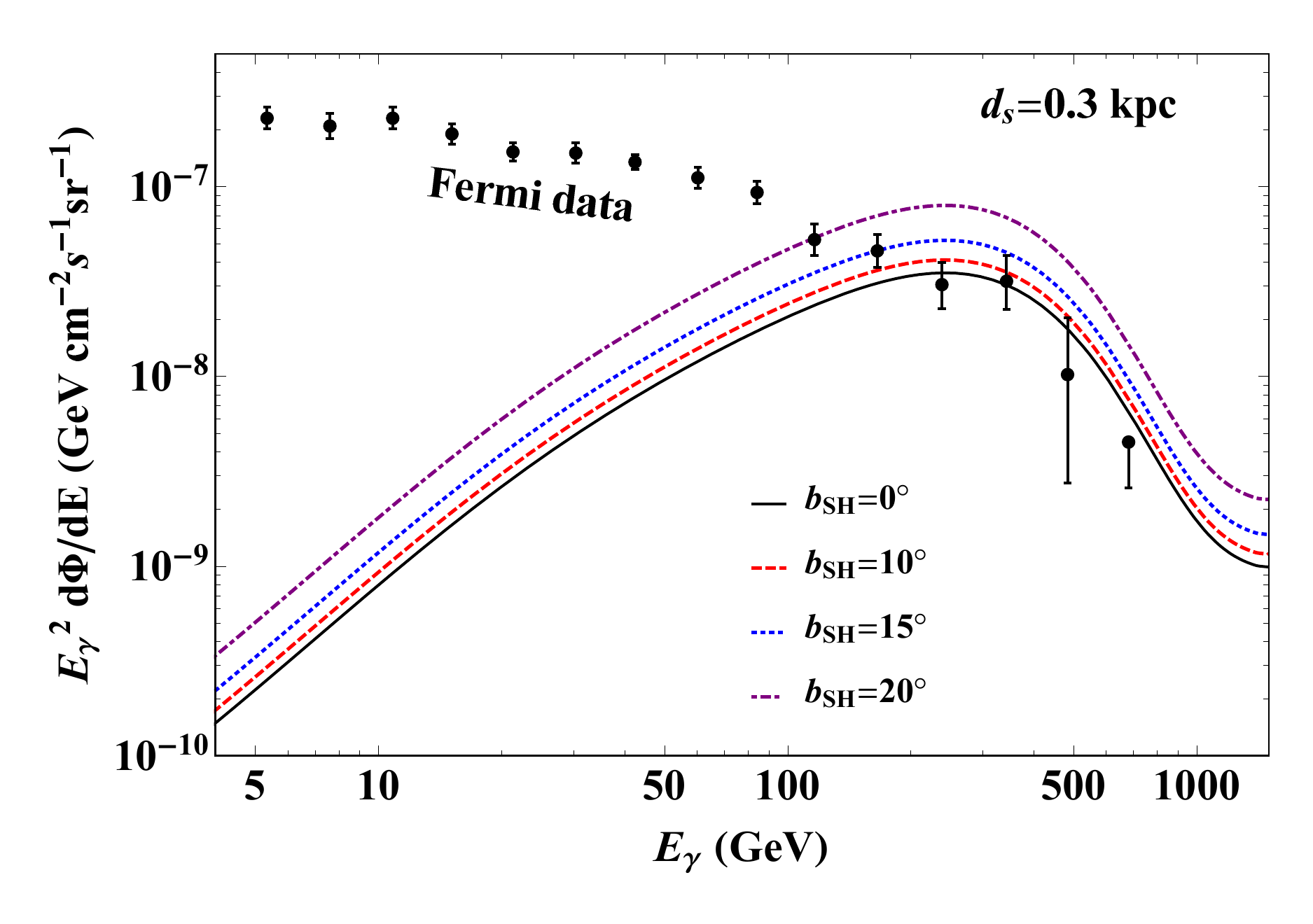}
\caption{
Left panel: Fermi IGRB data \cite{Ackermann:2014usa} 
and gamma rays from DM 
where $V_1$ is electrophilic 
and $V_2$ is $L_\mu-L_\tau$ gauged, 
with different $d_s$ values: $d_s=(0.1, 0.2, 0.3, 0.4)$ kpc. 
The subhalo is placed at $b_{\rm SH}=0^{\circ}$. 
Right panel: same as the left panel except that we keep  
$d_s=0.3$ kpc fixed and let $b_{\rm SH}$ vary: 
$b_{\rm SH} =(0^\circ, 10^\circ, 15^\circ, 20^\circ)$. 
}
\label{fig-fermi-iso-bg6}
\end{centering}
\end{figure}

Here we study the effects on the Fermi IGRB data by changing 
various parameters for the DM subhalo. 
The gamma ray flux is very sensitive to the distance 
between the subhalo and us. 
We compute the gamma rays expected at Fermi 
using different $d_s$ values on 
the left panel figure of Fig.\ (\ref{fig-fermi-iso-bg6}). 
Different $d_s$ values not only lead to different J-factors in the 
Fermi search region, but also lead to different DM 
annihilation cross sections which are provided in Table (\ref{tab-best-fits}), 
since one has to fit the DAMPE data. 
The predicted gamma rays become larger when the subhalo moves towards us. 
In order to evade the Fermi IGRB constraints, the subhalo has to be at least 
0.3 kpc away from us. 
We also compute the gamma rays from the subhalo when 
it moves away from the Galactic plane. 
The gamma ray flux expected in Fermi is shown on the 
right panel figure of Fig.\ (\ref{fig-fermi-iso-bg6}) for 
several different $b_{\rm SH}$ values. 
If the subhalo moves away from the Galactic plane for more than 
$10^\circ$, the gamma rays produced in the Fermi IGRB search region 
become significant above the current measurements.

\begin{figure}[!htbp]
\begin{centering}
\includegraphics[width=0.45\columnwidth]{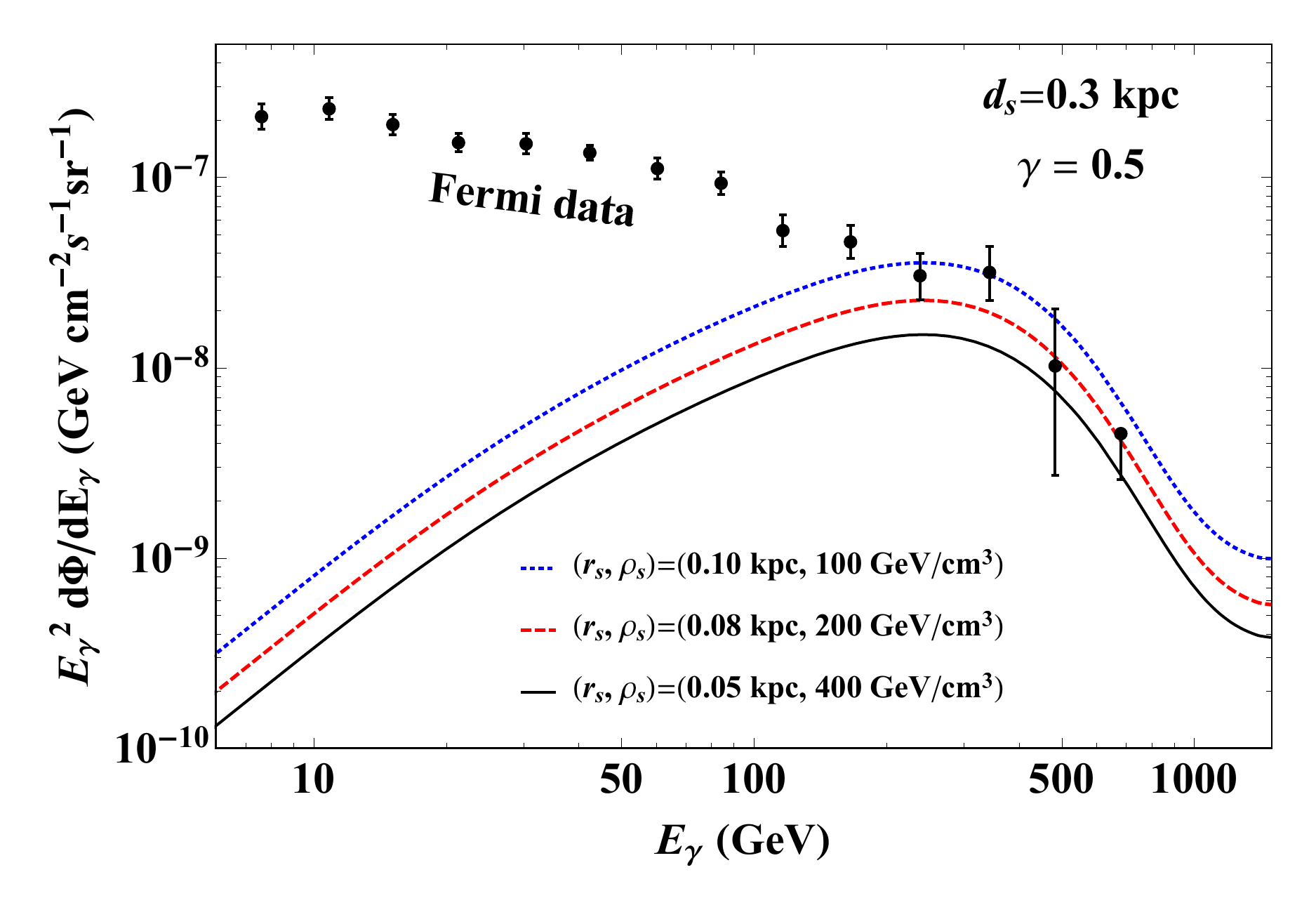}
\caption{Fermi IGRB data \cite{Ackermann:2014usa} 
and gamma rays from DM for different subhalo profiles. 
Here $V_1$ is electrophilic 
and $V_2$ is $L_\mu-L_\tau$ gauged.  
Here $d_s=0.3$ kpc and $b_{\rm SH} =0^\circ$.
	The DM annihilation cross sections are
	$\langle\sigma v\rangle_1=2.33\times10^{-26}$cm$^3$/s and 
	$\langle\sigma v\rangle_2=1.06\times10^{-24}$cm$^3$/s, 
	for the case where $r_s$=0.05 kpc and $\rho_s$=400 GeV/cm$^3$.
	The DM annihilation cross sections are
	$\langle\sigma v\rangle_1=2.26\times10^{-26}$cm$^3$/s and 
	$\langle\sigma v\rangle_2=1.05\times10^{-24}$cm$^3$/s, 
         for the case where $r_s$=0.08 kpc, $\rho_s$=200 GeV/cm$^3$. 
         For the case where $r_s$=0.1 kpc, $\rho_s$=100 GeV/cm$^3$, 
         the DM annihilation cross sections are
	listed in Table (\ref{tab-best-fits}).} 
\label{fig-fermi-rs-rhos}
\end{centering}
\end{figure}

We further study the gamma rays by changing the subhalo profile parameters 
$(r_s,\rho_s)$, in the Fig.\ (\ref{fig-fermi-rs-rhos}) where $d_s=0.3$ kpc 
and $\gamma=0.5$ are fixed. 
Two sets of parameters in addition to the default values for the subhalo 
are used here. 
For each case, the DM annihilation cross sections for the two 
different channels are chosen such that one obtains the least $\chi^2$ 
fit to the DAMPE data. 
As shown in Fig.\ (\ref{fig-fermi-rs-rhos}), the Fermi constraints can be 
significantly alleviated if the DM subhalo becomes smaller and denser.


\section{AMS constraints}
\label{sec:AMS}

We do not attempt to explain the AMS positron excess. 
However, the two-mediator DM model 
cannot produce too many positrons so that they violate the AMS 
data on the positron fraction measurement  \cite{Accardo:2014lma}.

\begin{figure}[!htbp]
\begin{centering}
\includegraphics[width=0.49\columnwidth]{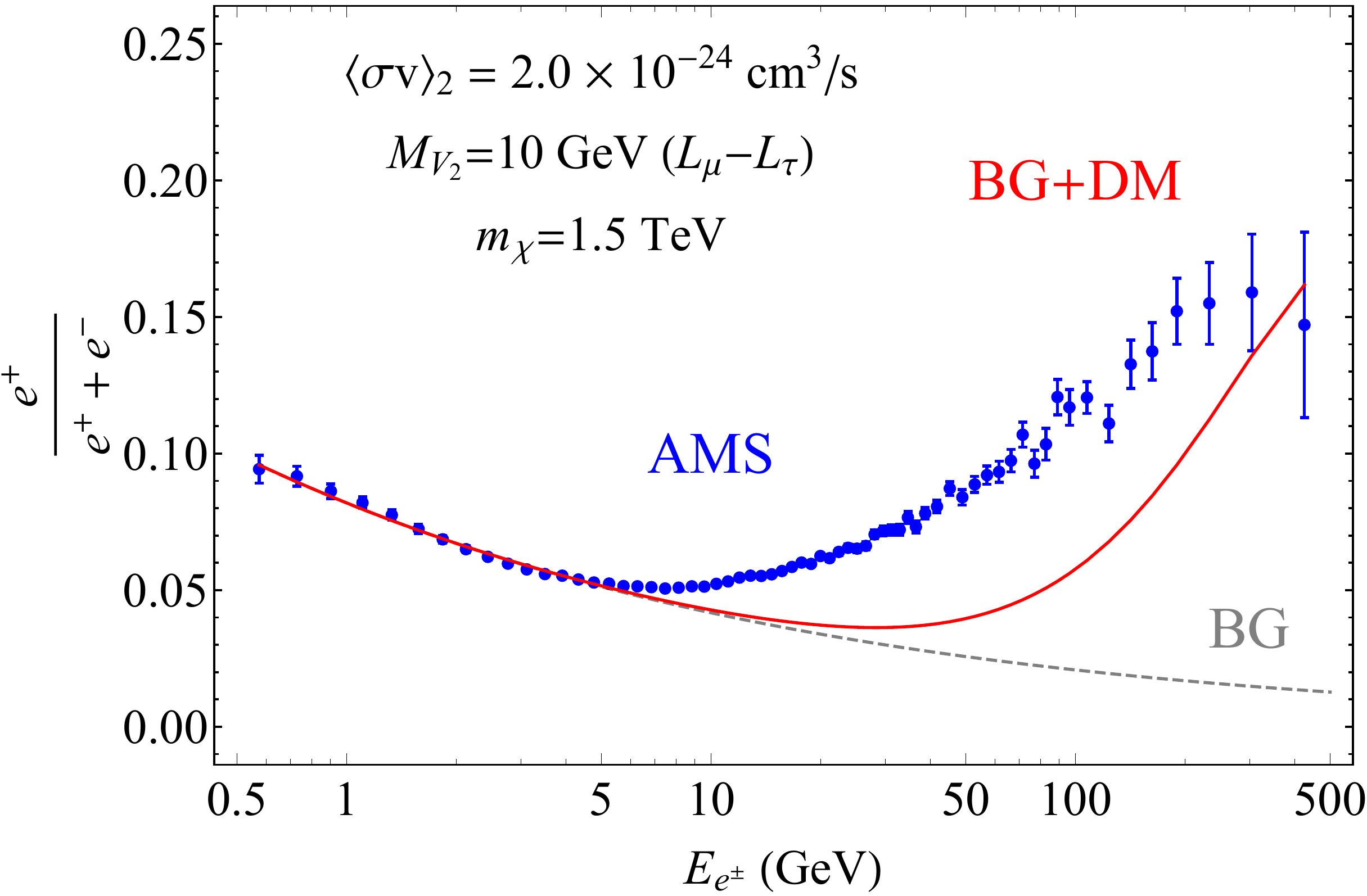}
\includegraphics[width=0.46\columnwidth]{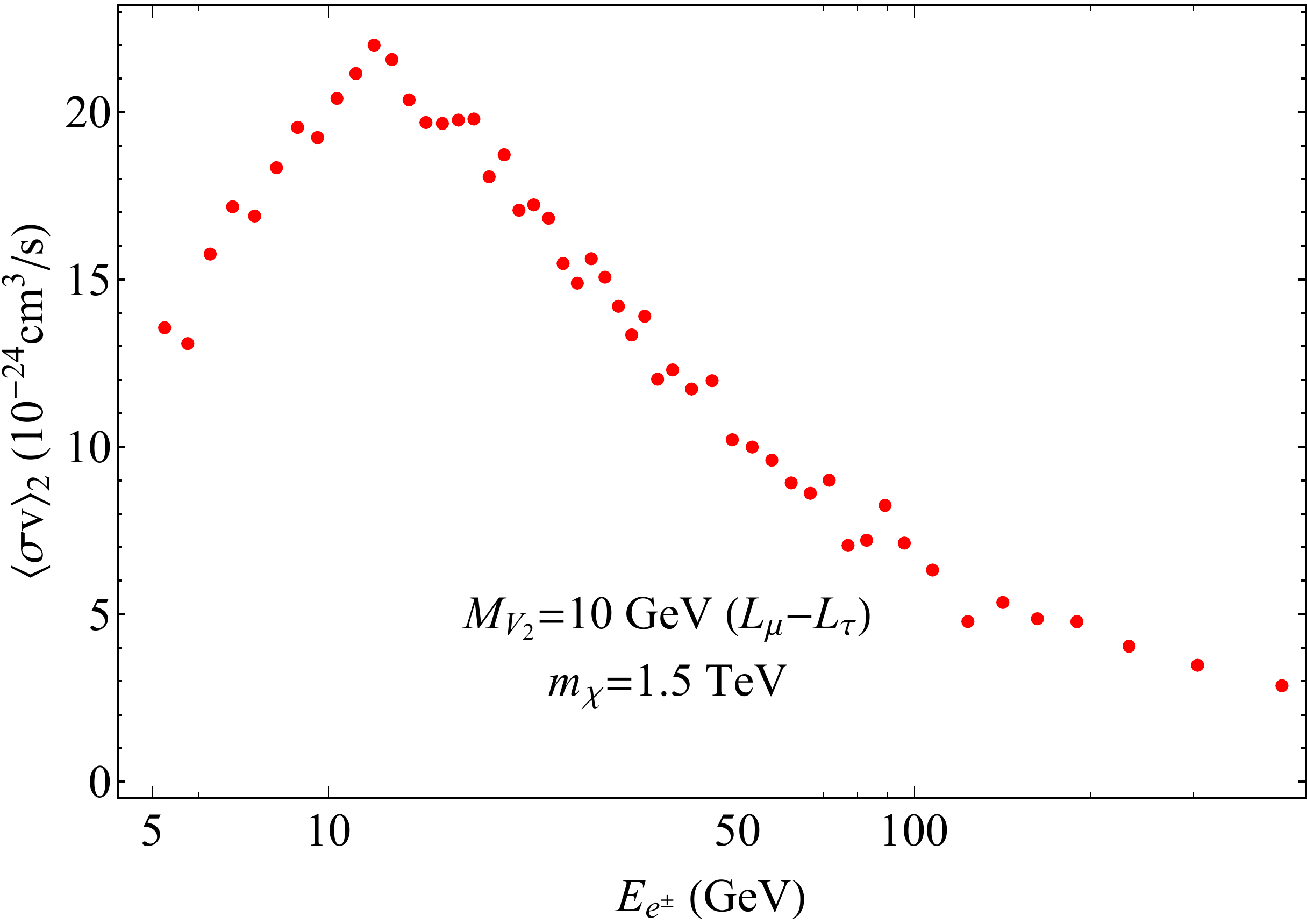}
\caption{Left panel: AMS positron fraction data \cite{Accardo:2014lma}
and the predicted value from the $L_\mu - L_\tau$ gauge boson. 
Here we use $m_\chi=1.5$ TeV, 
$m_{V_2}=10$ GeV, 
and 
$\langle \sigma v \rangle_{2}=  2.0 \times$ 10$^{-24}$ cm$^3$/s 
for the DM annihilation cross section. 
Right panel: 95$\%$ C.L. upper bound on $\langle \sigma v \rangle_{2}$
from each AMS data point. 
} 
\label{fig:AMS}
\end{centering}
\end{figure}

To compute the AMS constraints on the DM model, we 
extrapolate our simple cosmic ray electron/positron background given by Eq.\ (\ref{eq:bg}) 
down to low electron energy range. 
We further assume that the background of the positron fraction 
take the following simple expression $f^{\rm BG}=1/(C_{f} E^{\gamma_{f}} + 1)$. 
We use first 15 data points in the AMS positron fraction data \cite{Accardo:2014lma} 
to find the best-fit parameters: $C_{f} = 11.2$ and $\gamma_{f} = 0.31$. 
The positron fraction including contributions both from the background 
and from DM annihilations is thus computed by 
\be
f^{\rm th} = 
{\Phi^{\rm BG}f^{\rm BG}+\Phi^{\chi}/2 
\over 
\Phi^{\rm BG}+\Phi^{\chi}}
\ee
where $\Phi^{\chi}$ is the cosmic flux including both electron and positron 
due to DM annihilations. 
We use ($f_i^{\rm AMS}+1.64\, \delta f_i^{\rm AMS}$) 
at each AMS data point (excluding the first 15 points) 
to compute the 95\% C.L. upper bound on DM annihilation cross section, 
where $f_i^{\rm AMS}$ is the AMS positron fraction data 
and the $\delta f_i^{\rm AMS}$ is the error bar for each data point. 
Fig.\ (\ref{fig:AMS}) shows the AMS constraints on the DM annihilation cross section 
mediated by the $L_\mu - L_\tau$ gauge boson using the positron fraction data. 
The most stringent limit comes from the highest energy bin in the 
AMS data, which provides the 95\% CL upper bound as 
$\langle \sigma v \rangle_{2} \lesssim 3 \times 10^{-24}$ cm$^{3}$/s 
for the $L_\mu - L_\tau$ gauge boson. 
The predicted positron fraction values at the AMS energy range in our model lie 
below the AMS measurements. 
We note in passing that the gap between our predicted positron fraction 
and the actual AMS data could be due to astrophysical sources.

\section{LHC constraints}
\label{sec:ATLAS}

Here we study the LHC constraints on the $V_1$ boson that is kinetically mixed 
with the SM hypercharge. 
In this case, the $V_1$ boson couples to all SM fermions due to the kinetic mixing 
parameter $\epsilon$, which is given in Eq.\ (\ref{eq:eff_lag}). 
Thus, the $V_1$ boson can be produced in the Drell-Yan process 
at the LHC and can be searched for by reconstructing the dilepton final states. 
Here we utilize the recent ATLAS data \cite{Aaboud:2017buh} 
to put constraints on the 
kinetic mixing parameter $\epsilon$ between $V_1$ boson and the SM hypercharge boson.


\begin{figure}[!htbp]
\begin{centering}
\includegraphics[width=0.454\columnwidth]{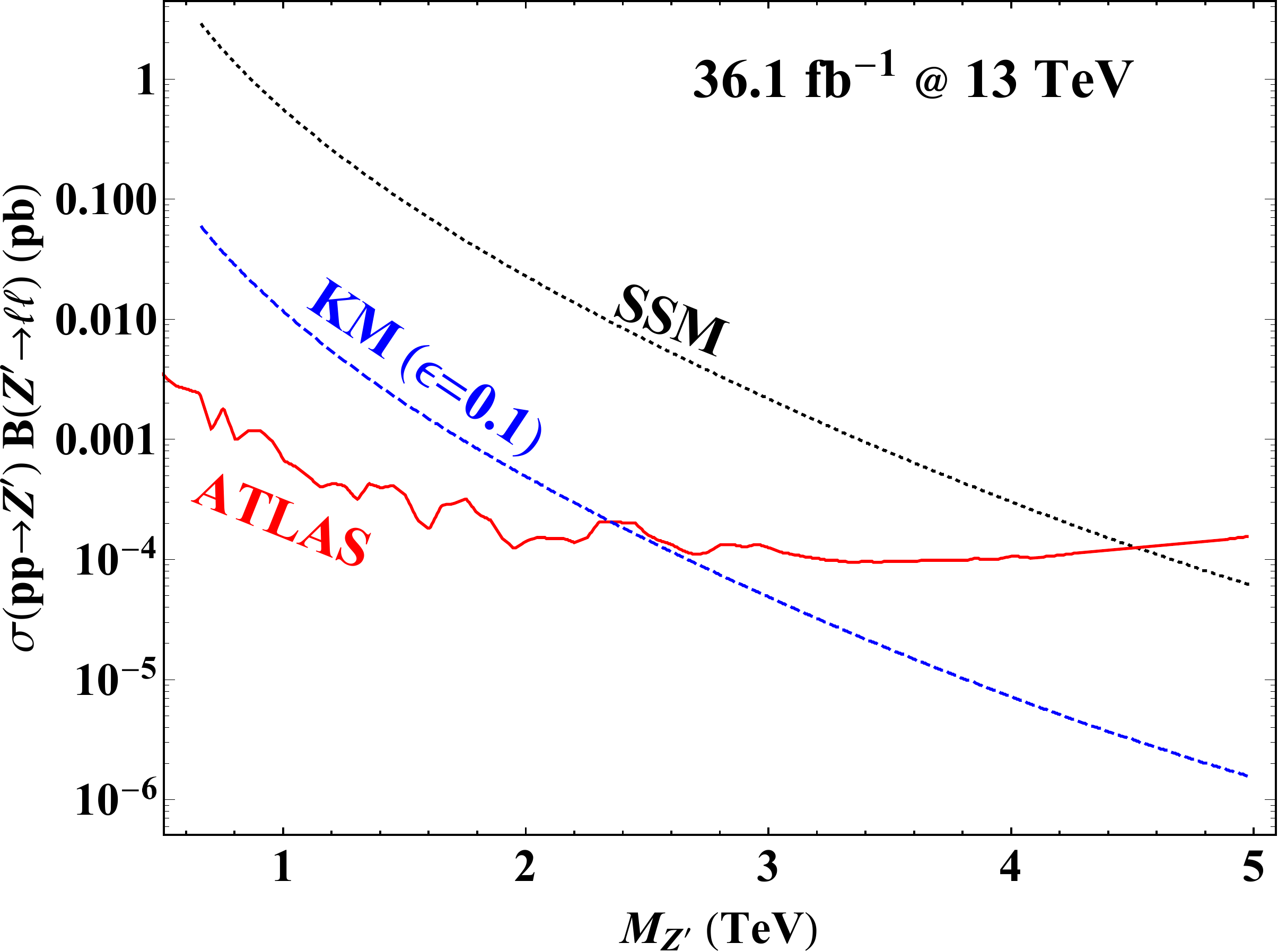}
\includegraphics[width=0.44\columnwidth]{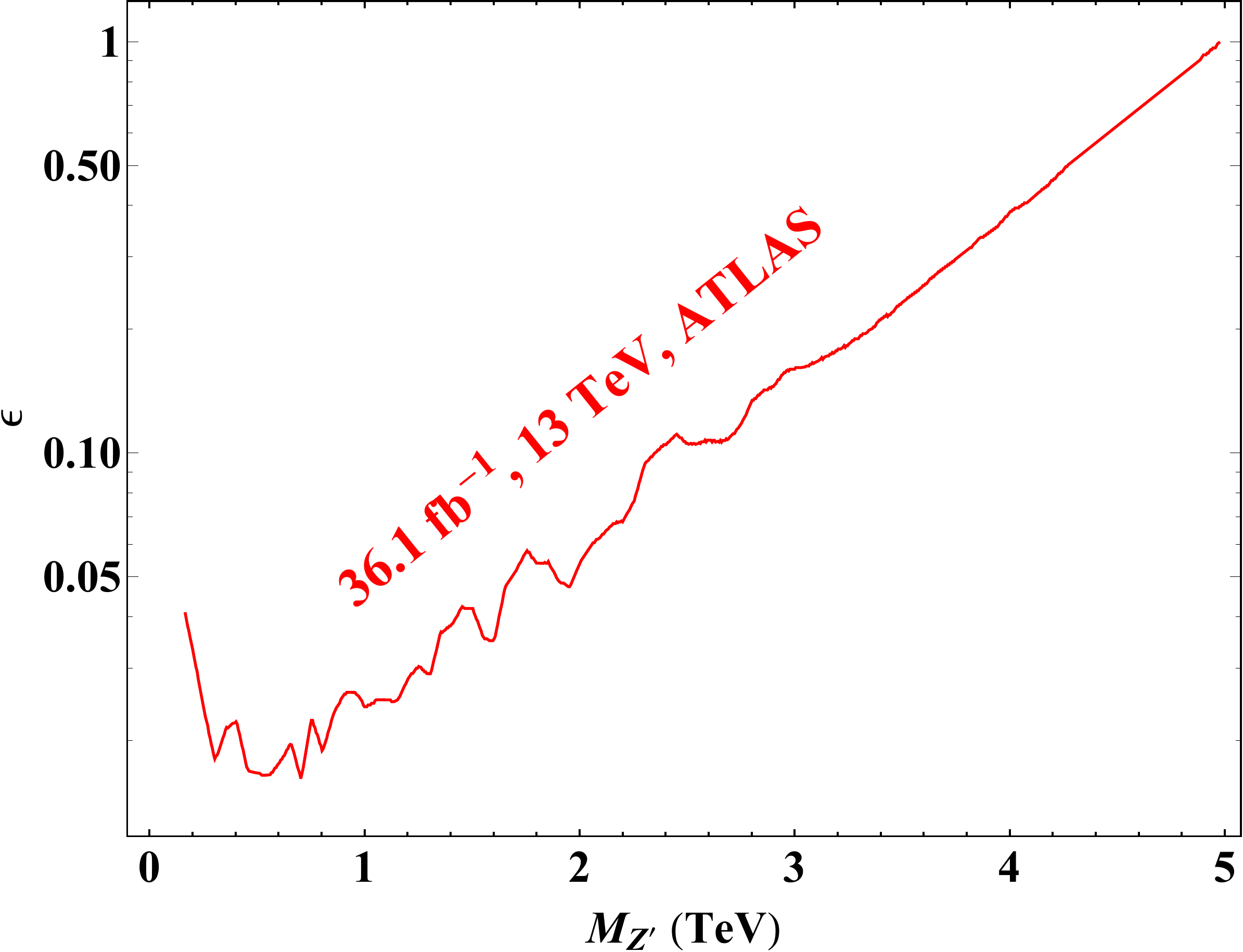}
\caption{Left panel: ATLAS constraints (13 TeV and 36.1 fb$^{-1}$) \cite{Aaboud:2017buh} 
on the $Z'$ boson in the dilepton channel. Overlaid are the 
predictions in the kinetic mixing (KM) model 
with parameter $\epsilon=0.1$ and in sequential SM (SSM) model. 
Right panel: ATLAS upper bound on $\epsilon$ in the KM model as 
a function of the $Z'$ mass.  
} 
\label{fig:atlas}
\end{centering}
\end{figure}

Fig.\ (\ref{fig:atlas}) shows the ATLAS upper bound on the dilepton production 
cross section, using 36.1 fb$^{-1}$ data at the 13 TeV colliding energy. 
Predicted dilepton signals arising from the kinetic-mixing model and from 
the sequential standard model are also shown on the left panel figure of Fig.\ (\ref{fig:atlas}). 
The dilepton cross section with $\epsilon=0.1$ for the 3 TeV $M_{Z'}$ boson 
in the kinetic-mixing model 
is below the current LHC limit. We further compute the upper bound 
on $\epsilon$ from the dilepton final states in 
the entire ATLAS search range, on the right panel figure 
of Fig.\ (\ref{fig:atlas}). 
The limit on $\epsilon$ will certainly improve when 
all data currently accumulated at the LHC are analyzed 
(about 150 fb$^{-1}$ data have been collected by ATLAS and by CMS individually so far \cite{zhang}). 
However, to reach the sensitivity of probing 
the model point considered in our analysis, 
$\epsilon=0.01$ for a 3 TeV $Z'$ boson, 
more data in future LHC runs are probably needed.


\section{Sommerfeld enhancement}
\label{sec:RD}

\begin{figure}[!htbp]
\begin{centering}
\includegraphics[width=0.65\columnwidth]{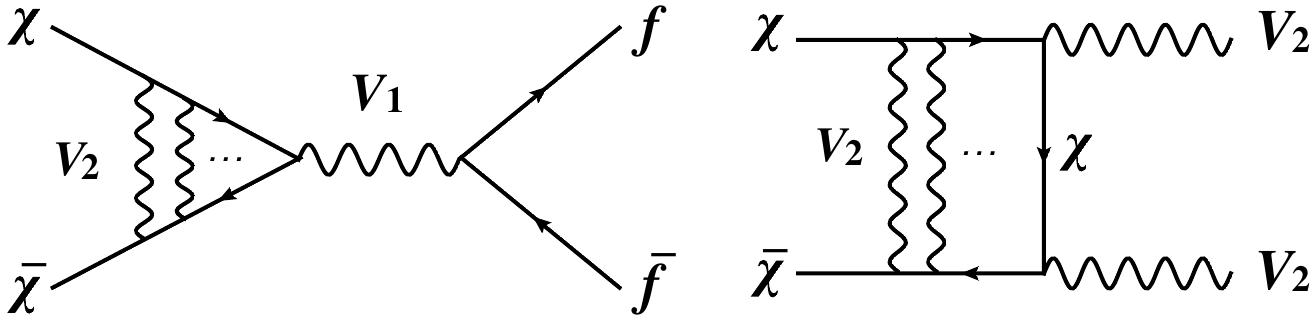}
\caption{Illustrations of light $V_2$ exchanges between annihilating 
DM particles in the two channels which induce the Sommerfeld enhancement.} 
\label{fig-SE}
\end{centering}
\end{figure}

The cross section of the process $\chi\chi \to V_2 V_2$ is larger than 
the canonical thermal DM annihilation cross section by about two orders of magnitude, 
which would suppress the DM abundance significantly.  
However, we should take into account the 
Sommerfeld enhancement induced via $V_2$ exchanges between 
DM particles in the annihilation processes as illustrated in Fig.\ (\ref{fig-SE}), 
since the mediator $V_2$ is light and the velocity of DM is low in the MW halo.

\begin{figure}[htbp]
\begin{centering}
\includegraphics[width=0.44\columnwidth]{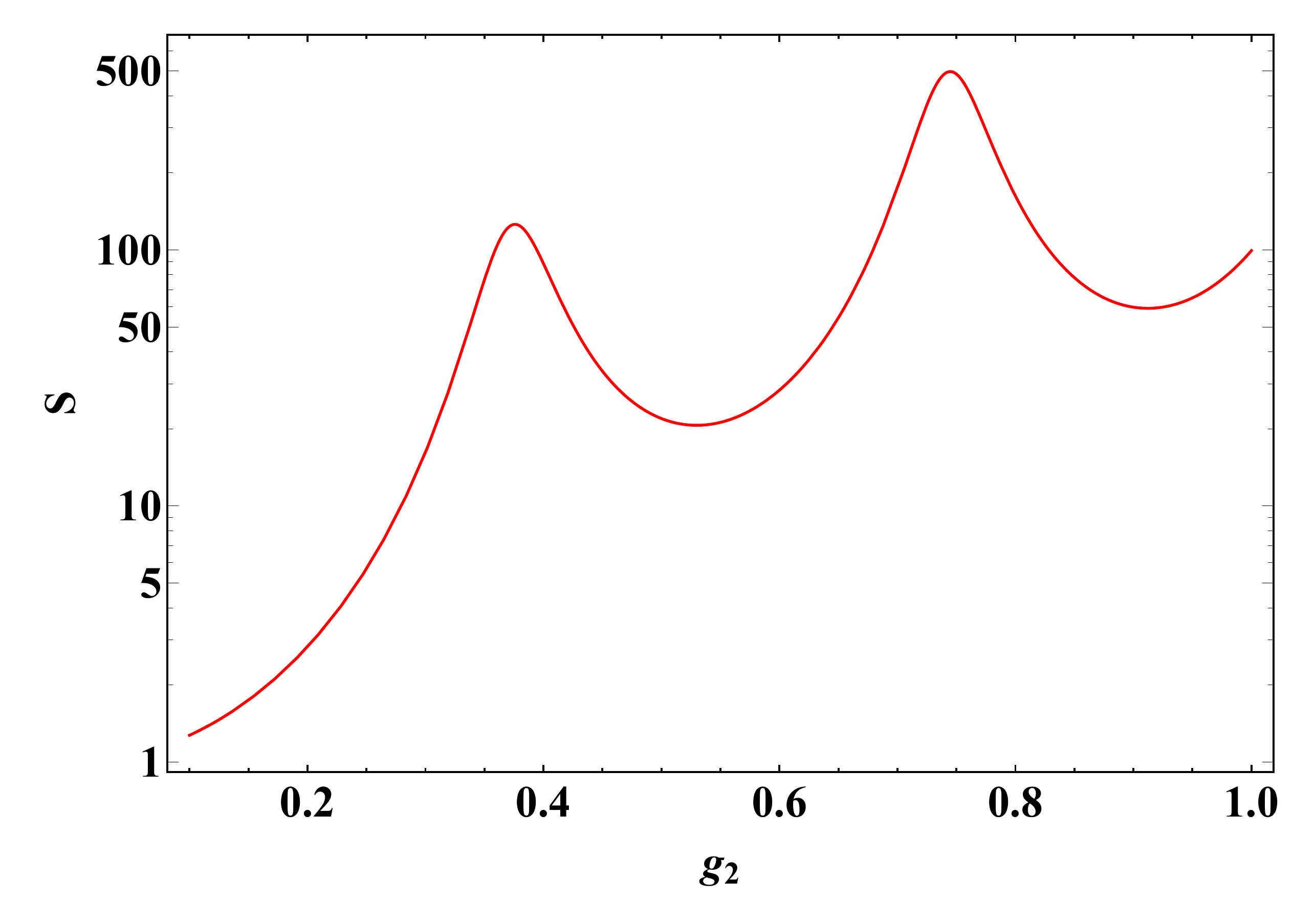}
\includegraphics[width=0.45\columnwidth]{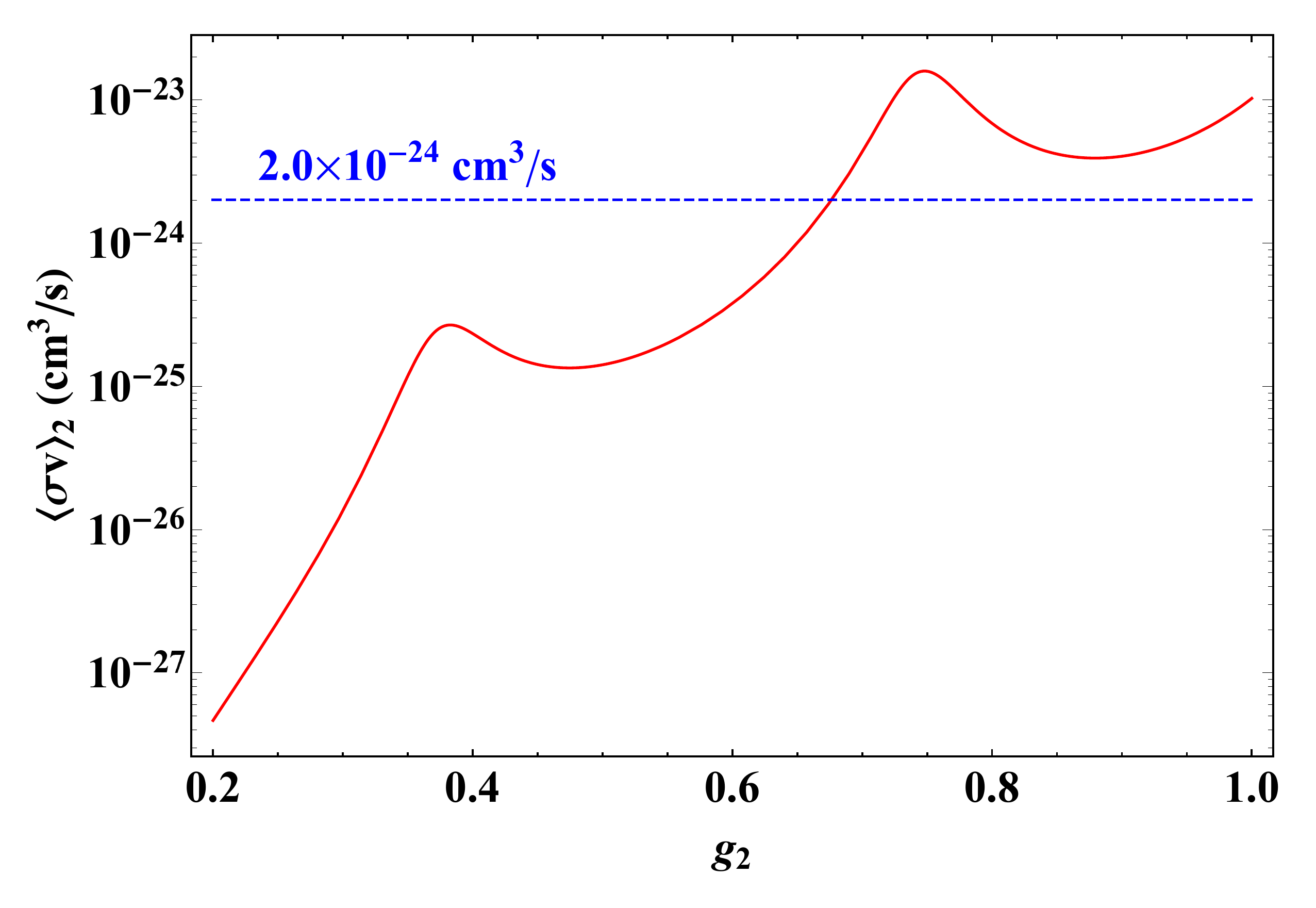}
\caption{Left panel: Sommerfeld enhancement factor $S$ as a function
of the coupling $g_2$ where $m_\chi=1.5$ TeV and $m_{V_2}=10$ GeV.  
Right panel: DM annihilation cross section $\langle \sigma v \rangle_2$ as a function
of $g_2$. The blue-dashed line indicates the needed cross section to fit DAMPE data. 
}
\label{fig:S}
\end{centering}
\end{figure}

The Sommerfeld enhancement factor $S$ can be approximated by 
\cite{Cassel:2009wt}
\cite{Slatyer:2009vg} 
\cite{Cline:2015qha} 
\be 
S = \left({\pi\over \epsilon_v}\right){\sinh
X \over \cosh X - \cos\sqrt{{(2\pi/\bar\epsilon_{2})} - X^2}}, 
\label{eq:sommerfeld} 
\ee 
where $\bar\epsilon_{2}= (\pi/12)\epsilon_{2}$ and $X
= \epsilon_v/\bar\epsilon_{2}$, 
and $\epsilon_{2} = {m_{V_2}/(\alpha_2 m_\chi)}$, 
$\epsilon_v = {v/\alpha_2}$ with $\alpha_2=g_2^2/(4\pi)$. 
We take $v=10^{-3}$ as the typical DM velocity in the halo. 
The left panel figure of Fig.\ (\ref{fig:S}) shows the 
Sommerfeld enhancement factor as a function of 
the gauge coupling $g_2$ where 
the mediator $V_2$ mass is 10 GeV and the dark matter 
mass is 1.5 TeV.

For the process $\chi\chi\to V_2 V_2$, the DM annihilation cross section is given by 
$\langle\sigma v\rangle_2 = \langle\sigma v\rangle_2^0  \times S(g_2)$ 
where $\langle\sigma v\rangle_2^0  \simeq {g_2^4/(16 \pi m_\chi^2)} $ is the annihilation cross section
without taking account the Sommerfeld enhancement effect. 
By equating this expression with $2.0\times10^{-24}$ cm$^3$/s, 
the needed cross section to fit DAMPE, 
one obtains $g_2 =0.68$. Thus one obtains the corresponding
Sommerfeld enhancement factor  $S \simeq 93$. 
We further plot $\langle\sigma v \rangle_2$ as a function of  coupling $g_2$ in
the right panel figure of  Fig.\ (\ref{fig:S}).

For the  process $\chi\chi\to V_1 \to f \bar{f}$, one also has to consider 
the same enhancement due to the $V_2$ mediator, so that the DM annihilation cross section 
should be computed via $\langle\sigma v\rangle_1=S \times\langle\sigma v\rangle_1^0$, 
where the superscript $0$ indicates the cross section without taking the Sommerfeld 
enhancement into account. 
Using $S\simeq 93$, we find that 
the model point ($\epsilon$, $g_1$, $m_{\chi}$, $M_{V_1}$) 
= (0.01, 0.1, 1500 GeV, 2994.2 GeV) 
in the parameter space of the KM model can give rise to 
$\langle\sigma v\rangle_1 = 3.9\times10^{-25}{\rm cm^3/s}$ 
which is needed to fit the DAMPE data.

The DM relic abundance which is primarily determined by the 
DM annihilation cross section at the so-called freeze-out epoch at the early universe. 
Typical freeze-out occurs at the temperature $T\simeq m_\chi /(20-25)$ such that  
the DM velocity is approximately $v\simeq1/4$, 
where DM annihilation cross section no longer receives significant Sommerfeld 
enhancement that is present at the current galaxy. 
We compute the DM annihilation cross section for the 
processes $\chi\chi\to V_1\to f\bar{f}$ (KM) 
and $\chi\chi\to V_2V_2$ at 
the freeze-out and find that 
$\langle\sigma v\rangle_1 = 1.0\times 10^{-28}$ cm$^3$/s, 
and $\langle\sigma v\rangle_2 =2.2\times 10^{-26}$ cm$^3$/s, 
when $T = m_\chi /25$. 
Thus the total DM annihilation cross section is approximately 
$2.2\times 10^{-26}$ cm$^3$/s at freeze-out which is very close to the 
canonical thermal DM annihilation cross section needed 
to generate the right DM relic density in the universe. 
We note that there is a 
three orders of magnitude 
boost on $\langle\sigma v\rangle_1$ at 
current galaxy relative to the early universe,  
owing to both the Breit-Wigner enhancement and the 
Sommerfeld enhancement in this annihilation channel.

\section{Conclusions}
\label{sec:sum}

There are two exotic features present in the new cosmic electron spectrum 
observed by the DAMPE collaboration, 
including a break at 0.9 TeV and a peak at 1.4 TeV. 
We propose  
to simultaneously explain both features in the DAMPE data via annihilations from 
one DM species that interacts with SM via two different mediators. 
Thus two different DM annihilations channels via the two 
mediators generate the two new features in the 
cosmic electron energy spectrum near TeV. 
The annihilation process mediated by the heavier $V_1$ boson 
generates the 1.4 TeV peak; 
the annihilation process mediated by the lighter $V_2$ boson 
produces the extended break near 0.9 TeV.

In this work, we consider two concrete examples of the two-mediator DM models. 
In both cases the lighter $V_2$ boson is 
$L_\mu-L_\tau$ gauged and has mass 10 GeV such that 
$V_2$ can be on-shell produced in annihilations of DM  
which is taken to be 1.5 TeV. 
We consider the heavier $V_1$ boson to be either 
electrophilic or kinetically mixed with the SM hypercharge.

We assume a single power-law cosmic electron background 
which contains only two parameters and a DM subhalo which is 0.3 kpc from us. 
Both electrophilic and KM $V_1$ bosons provide 
good fits to the 1.4 TeV excess, 
with the annihilation cross section  
$4.9\times10^{-26}$ cm$^3$/s and 
3.9 $\times$ 10$^{-25}$ cm$^3$/s respectively; 
the $L_\mu-L_\tau$ gauge boson $V_2$ provides a 
good fit to the break 
with the annihilation cross section 
2.0 $\times$ 10$^{-24}$ cm$^3$/s.

Several experimental constraints on the DM models are analyzed, 
including HESS, 
Fermi IGBG, 
AMS positron fraction and LHC dilepton searches. 
Gamma rays expected at the HESS search region 
are mainly coming from annihilations via the $V_2$ boson 
due to the larger cross section. 
HESS constraints are very sensitive to the DM density profile 
for the MW halo. 
The needed cross section for the $V_2$ process 
is excluded if one assumes the NFW or Einasto profile for the MW halo, 
but still allowed if the isothermal profile is considered. 
In addition, a substantial amount of gamma rays also arise in DM 
annihilations via the kinetic-mixing $V_1$ boson; 
we find that the gamma rays from both annihilation channels 
are consistent with HESS data assuming the isothermal profile for the MW halo. 
We also find that the subhalo cannot be put at the Galactic center direction 
since it would contribute a significant amount of gamma rays 
to the HESS search region. 
Fermi isotropic gamma ray background constraints are sensitive 
to the distance between the subhalo and us. 
We find that our models do not violate the Fermi isotropic gamma ray 
background if the subhalo is placed at 0.3 kpc from us. 
We also note that one can 
begin to probe our model with more data accumulated at Fermi.  
DM annihilations in our model cannot provide satisfactory explanations 
to the AMS positron fraction excess. Nonetheless, one can use the 
AMS data to put the constraints on DM models by demanding that the 
predicted positron fraction in DM models not exceed the AMS measurement. 
We find that the highest energy bin in the AMS data gives the 
most stringent bound on the $L_\mu-L_\tau$ gauge boson process, 
and will probe our model in the near future. 
LHC constraints on the KM $V_1$ boson are analyzed in the 
dilepton channel. For a 3 TeV $V_1$ boson, the upper bound on 
$\epsilon$ is about 0.1.

The DM annihilation cross sections needed to fit DAMPE data 
are much larger than the canonical thermal cross section. 
This discrepancy can be nicely explained by the Sommerfeld enhancement 
due to the light $V_2$ mediator in the models.  
Taking into account the non-perturbative Sommerfeld enhancement corrections 
present in the current galaxy, our model is consistent with the relic density 
requirement in the thermal DM framework.


\acknowledgments


We thank Farinaldo Queiroz and Lei Zhang for helpful correspondence. 
The work is supported in part  
by the National Natural Science Foundation of China under Grant Nos.\ 
11775109 and U1738134, 
by the National Recruitment Program for Young Professionals, 
by the Nanjing University Grant 14902303, 
by the National Postdoctoral Program for Innovative Talents
under Grant No.\ BX201700116, 
and by the Jiangsu Planned Projects for Postdoctoral Research Funds 
under Grant No.\ 1701130B.

\appendix


\section{HESS J-factors}
\label{app:Jfactor}

Here we compute the HESS J-factor for different DM profiles. 
The HESS signal region is 
a circular region of $1^{\circ}$ radius 
excluding a $\pm 0.3^{\circ}$ band in Galactic latitudes \cite{Abdallah:2016ygi}.
We consider three different DM density profiles (NFW, Einasto, and Isothermal) 
which have been used in HESS analysis. 
The NFW profile is given by 
\be
\rho_{\rm NFW} (r) =  { \rho_s \over (r/r_s)(1+(r/r_s)^2}, 
\ee
where we use $r_s = 21$ kpc \cite{Pieri:2009je}. 
The Einasto profile is given by 
\be
\rho_{\rm E} (r) =  \rho_s \exp 
\left[-\frac{2}{\alpha}\left( \left(\frac{r}{r_s} \right)^\alpha-1 \right) \right],
\ee
where we use $\alpha = 0.17$ and $r_s = 20$ kpc \cite{Pieri:2009je}.                     
The isothermal profile is given by 
\be
\label{eq-isothermal}	
\rho_{\rm iso} (r) =  \frac{\rho_s}{(r/R)^\gamma (1+(r/R)^\alpha)^{(\beta-\gamma)/ \alpha}},
\ee
where we use $R = 3.5$ kpc, $\alpha = 2$, $\beta = 2$ and $\gamma = 0$ 
\cite{Bertone:2004pz}. The value of 
$\rho_s$ in all the above profiles are chosen such that the 
local DM density is normalized to  
$\rho_\chi (8.5~{\rm kpc})  = 0.39 $ GeV/cm$^3$. 
We compute the HESS J-factors for these DM profiles using Eq.\ (\ref{eq:Jfactor})
\bea
J_{\rm NFW} &=& 2.25 \times 10^{21} ~ {\rm GeV^2/cm^5},        \nonumber \\         
J_{\rm E} &=& 4.41 \times 10^{21} ~ {\rm GeV^2/cm^5},            \nonumber \\          
J_{\rm iso} &=& 7.23 \times 10^{19}  ~ {\rm GeV^2/cm^5}. 
\eea
HESS collaboration \cite{Abdallah:2016ygi} provides the J-factors for two profiles: 
$J_{\rm NFW} = 2.67 \times 10^{21} ~ {\rm GeV^2/cm^5}$ 
and $J_{\rm E} = 4.92 \times 10^{21} ~ {\rm GeV^2/cm^5}$. 
Thus our calculation here yields slightly smaller J-factors than HESS. 
We use the $J_{\rm NFW}$ and $J_{\rm E}$ values 
provided by HESS \cite{Abdallah:2016ygi} 
in the rescaling method to be described in Appendix (\ref{HESSlimit}). 
Because the J-factor for the isothermal profile is not given explicitly 
by HESS \cite{Abdallah:2016ygi}, 
we use our calculated  $J_{\rm iso}$ value in the analysis.

\section{Rescaling method for HESS limits} 
\label{HESSlimit}

\begin{figure}[htbp]
\begin{centering}
\includegraphics[width=0.45\textwidth]{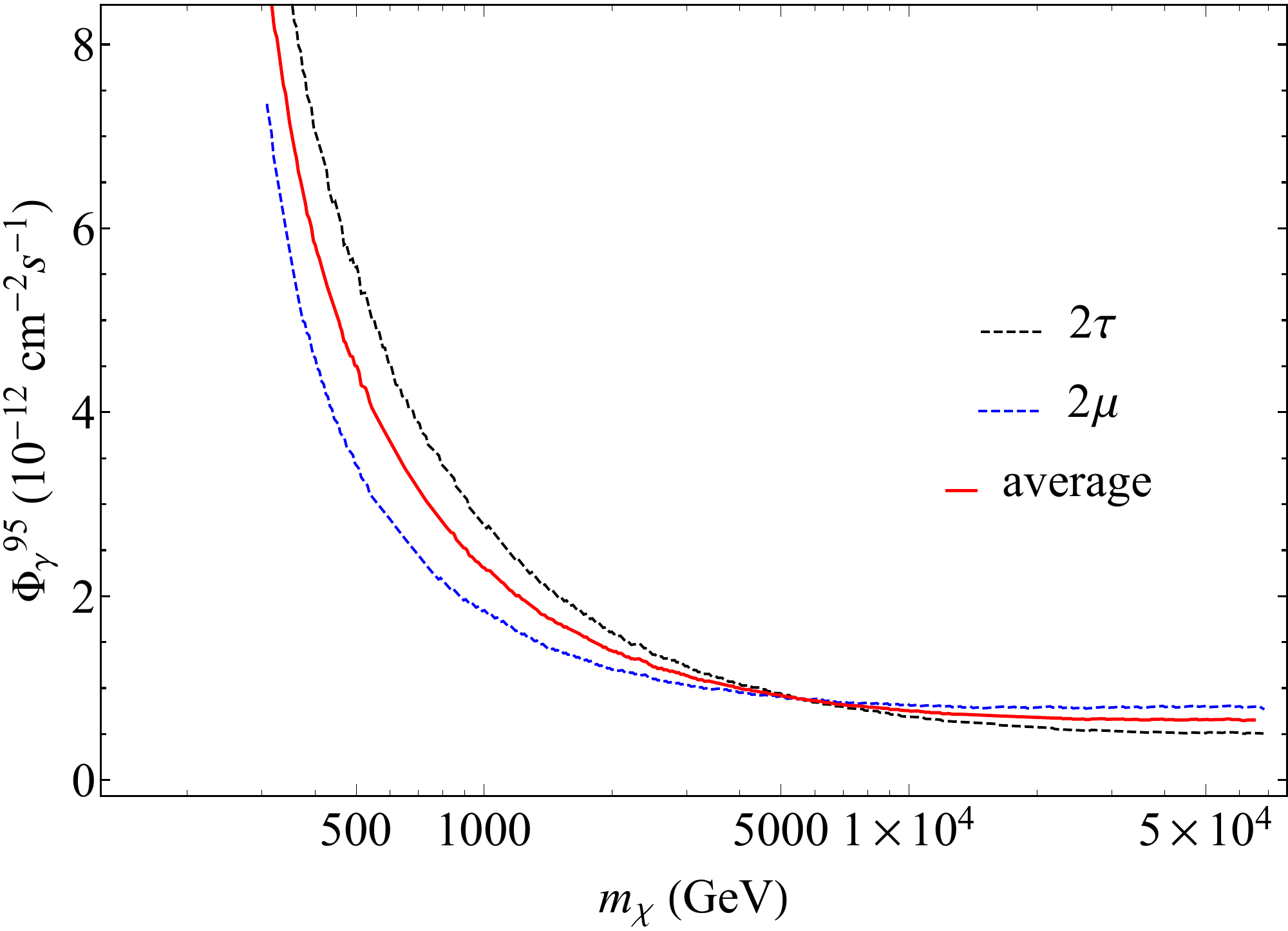}
\caption{HESS 95\% CL upper bound on the total gamma ray 
flux $\Phi_\gamma^{95}$ in the energy range 
$160$ GeV $<E_\gamma<m_\chi$  \cite{Abdallah:2016ygi}. 
The $2\mu$ ($2\tau$) curve is obtained based on the 
HESS upper bound on the annihilation cross section curve in the 
$\chi \chi \to \mu^+ \mu^-$ ($\chi \chi \to \tau^+ \tau^-$) 
channel \cite{Abdallah:2016ygi}. 
The ``average'' curve is the arithmetic mean of the 
$2\mu$ and $2\tau$ curves. 
}
\label{fig:phi95}
\end{centering}
\end{figure}

Here we describe the ``rescaling'' method used in our analysis to obtain the HESS limits. 
We first digitize the 95\% CL upper bound on DM annihilation cross section 
 $\langle \sigma v \rangle^{95}(m_\chi)$  from the 
HESS 254-h data analysis \cite{Abdallah:2016ygi} for one specific annihilation channel, 
for instance the $\chi\chi \to \mu^+ \mu^-$ annihilation channel. 
The 95\% CL upper bound on the total gamma ray flux $\Phi_\gamma^{95}(m_\chi)$ 
can be obtained by 
integrating the differential flux given in Eq.\ (\ref{eq:gammaflux}) over the 
gamma ray energy range $160$ GeV $<E_\gamma<m_\chi$, 
where we used the Einasto J-factor $J_{\rm E}$ 
given in Ref.\ \cite{Abdallah:2016ygi}. 
The HESS constraints can then be calculated for different DM annihilation channels 
by integrating the differential flux given in Eq.\ (\ref{eq:gammaflux}) 
taking into account different halo profiles and different gamma ray energy spectra. 
Fig.\ (\ref{fig:phi95}) shows the HESS 95\% CL bound on $\Phi_\gamma^{95}(m_\chi)$ 
for the $\chi \chi \to \mu^+ \mu^-$ and $\chi \chi \to \tau^+ \tau^-$ 
annihilation channels. 
For the on-shell produced vector bosons in the $L_{\mu} - L_{\tau}$ model, 
we use the average value of $\Phi_\gamma^{95}(m_\chi)$ of the 
$\chi \chi \to \mu^+ \mu^-$ 
and 
$\chi \chi \to \tau^+ \tau^-$ 
annihilation channels to compute the HESS limits.

\end{document}